\documentclass[conference,compsoc]{IEEEtran}
\usepackage{tikz}
\usepackage{amsmath}

\usepackage{times,url,color,soul,xspace,enumitem}
\usepackage{graphicx}
\usepackage{caption}
\usepackage{float}  
\usepackage{subcaption}
\usepackage{comment}
\usepackage{xspace}
\usepackage{tikz}
\usepackage{amsmath}
\usepackage[inline,draft,nomargin,index]{fixme}
\usepackage{hyperref}  

\usepackage{cleveref}
\crefformat{section}{\S#2#1#3} 
\crefformat{subsection}{\S#2#1#3}
\crefformat{subsubsection}{\S#2#1#3}
\usepackage{amsmath}
\usepackage{graphicx}  
\usepackage{xcolor}        
\usepackage[linesnumbered,ruled,vlined]{algorithm2e}
\usepackage{caption}
\usepackage{float}
\usepackage{graphicx}  
\usepackage{tabularx}
\usepackage{stfloats}
\usepackage{placeins}
\usepackage{listings}
\usepackage{float}  
\usepackage{xcolor}
\usepackage{tikz}
\usepackage{booktabs}
\newcommand*\circled[1]{\tikz[baseline=(char.base)]{
            \node[shape=circle,draw,inner sep=1pt] (char) {#1};}}

\definecolor{kw}{RGB}{0,0,255}    
\definecolor{cm}{RGB}{34,139,34}  
\definecolor{st}{RGB}{255,69,0}   

\lstdefinestyle{csecurity}{
  language=C,
  basicstyle=\ttfamily\scriptsize,
  numbers=left,
  numberstyle=\tiny\color{black},
  stepnumber=1,
  numbersep=8pt,
  showstringspaces=false,
  breaklines=true,
  tabsize=2,
  keywordstyle=\color{kw}\bfseries,  
  commentstyle=\color{cm}\itshape,   
  stringstyle=\color{st},            
  backgroundcolor=\color{white},     
  frame=none,
  framerule=0.3pt,
  xleftmargin=1.2em,
  framexleftmargin=1.0em,
  keepspaces=true,
  escapechar=§ 
}

\lstdefinestyle{asmgas}{
  basicstyle=\ttfamily\scriptsize,
  keywordstyle=\color{blue}\bfseries,
  commentstyle=\color{teal!70!black}\itshape,
  numbers=left,
  numberstyle=\tiny,
  stepnumber=1,
  numbersep=-5pt,
  tabsize=2,
  showstringspaces=false,
  breaklines=true,
  columns=fullflexible,
  captionpos=t,
  morekeywords={mov, movl, lea, push, call, rep, movsl, add, sub, cmp, jle, cmpl, ret}
}

\newcommand{\hlc}[1]{\textbf{\color{red}#1}}
\newcommand{\ubw}[1]{\colorbox{yellow}{\strut #1}} 

\SetCommentSty{mycommfont}

\newcommand{\algcomment}[1]{\textcolor{blue}{\scriptsize $\triangleright$ #1}}

\SetAlgoNlRelativeSize{-1}

\DontPrintSemicolon  

\FXRegisterAuthor{gt}{gtan}{\textcolor{orange}{GT}}

\FXRegisterAuthor{ms}{msantra}{\textcolor{magenta}{MS}}

\FXRegisterAuthor{dz}{dzeng}{\textcolor{orange}{DZ}}

\ifCLASSOPTIONcompsoc
  \usepackage[nocompress]{cite}
\else
  \usepackage{cite}
\fi
\ifCLASSINFOpdf
\else
\fi
\begin{document}


%

\date{}

\newcommand{\ida}{IDA~Pro}
\newcommand{\ghidra}{Ghidra}
\newcommand{\sysname}{iResolveX} 
\newcommand{\bpa}{BPA}
\newcommand{\callee}{CALLEE}
\newcommand{\attncall}{AttnCall}
\newcommand{\schedex}{SchedExec}
\newcommand{\layervsa}{BPA}           
\newcommand{\layerScoreGen}{iScoreGen}   
\newcommand{\layerScoreRefine}{iScoreRefine} 
\newcommand{\layerScoreTune}{iScoreTune} 
\newcommand{\pIndGraph}{p-IndirectCFG}          
\newcommand{\todo}[1]{\textbf{\textcolor{red}{#1}}}

\title{\Large \bf \sysname{}: Multi-Layered Indirect Call Resolution via Static Reasoning and Learning-Augmented Refinement}


\author{%
\IEEEauthorblockN{Monika Santra\IEEEauthorrefmark{1},
Bokai Zhang\IEEEauthorrefmark{1},
Mark Lim\IEEEauthorrefmark{2},
Vishnu Asutosh Dasu\IEEEauthorrefmark{1},
Dongrui Zeng\IEEEauthorrefmark{2},
Gang Tan\IEEEauthorrefmark{1}}
\IEEEauthorblockA{\IEEEauthorrefmark{1}The Pennsylvania State University\\
\{mvs7165,bzz5205,vdasu,gtan\}@psu.edu}
\IEEEauthorblockA{\IEEEauthorrefmark{2}Palo Alto Networks, Inc.\\
\{malim,dzeng\}@paloaltonetworks.com}
}


\maketitle


%
\IEEEpeerreviewmaketitle

\begin{abstract}
Indirect call resolution remains a key challenge in reverse engineering and control-flow graph recovery, especially for stripped or optimized binaries. Static analysis guarantees soundness but over-approximates, leading to many false positives, whereas machine-learning methods improve precision but sacrifice completeness and generalization. Existing tools like Ghidra, IDA, and Angr lag behind academic approaches in terms of efficacy.
We present \sysname{}, a hybrid multi-layered framework that combines conservative static analysis with learning-based refinement. The first layer uses a conservative value-set analysis (BPA) to ensure high recall. The second layer incorporates a learning-based soft signature scorer (iScoreGen) and selective inter-procedural backward analysis with memory inspection (iScoreRefine) to reduce false positives. The final output, \pIndGraph, annotates indirect edges with confidence scores, making it adaptable for diverse downstream analyses that demand different precision–recall trade-offs. \sysname{}’s iScoreGen reduces predicted targets by an average of 19.2\% across SPEC2006 and real-world binaries while maintaining BPA-level recall at 98.2\%. When combined with iScoreRefine, the total reduction reaches 44.3\% over BPA with a 97.8\% recall, representing only a 0.4\% loss. Furthermore, \sysname{} adapts to diverse analysis needs—offering conservative, recall-preserving or F1-optimized configurations—while outperforming state-of-the-art systems across both settings.

\end{abstract}

\section{Introduction}



Binary analysis, also known as binary reverse engineering (RE), is crucial to contemporary software security practices. It supports vital functions like vulnerability detection, malware classification, exploit triage, and the overall understanding of programs. Central to these operations is the control-flow graph (CFG), which effectively models the execution paths within a binary. The development of an accurate and comprehensive CFG~\cite{cfg_recovery} is pivotal for various downstream applications, including control-flow integrity (CFI)~\cite{cfi1, cfi2, cfi3, cfi_ano}, symbolic execution~\cite{dirsym, fuzzsym}, decompilation~\cite{decomp}, taint tracking~\cite{taint}, and fuzz testing~\cite{fuzz2, symexectest}. Inaccuracies within the CFG, resulting from missing edges or imprecise execution paths, can lead to overlooked vulnerabilities, misleading findings, and inadequate test coverage. Additionally, as Graph Neural Networks (GNNs)~\cite{gnn_cfg_2, gnn_using_cfg1} and Large Language Models (LLMs)~\cite{llm_cfg_1, llm_cfg_2} increasingly rely on CFG-derived representations, the accuracy of these graphs becomes crucial for the supervision signal. Errors in CFG recovery can propagate as biased training data, which undermines correctness and reliability. 
A significant challenge in recovering CFGs is the resolution of indirect calls~\cite{bpa}, whose targets are determined at runtime via registers, memory, or computed addresses. This introduces significant uncertainty, complicating static analysis. Additionally, dynamic or hybrid analysis~\cite{SchedExec, dyn1, dyn2} often faces limitations related to coverage, and the issue becomes more pronounced when dealing with stripped or optimized binaries that lack crucial symbol and type information. To address this challenge, research efforts have largely followed two parallel paths---\textit{static analysis}~\cite{analysis1, bindsa, bpa, SchedExec, analysis2, analysis3, analysis4} and \textit{ML-based approaches}~\cite{callee, Attncall}---each offering complementary strengths and limitations. 

Static analyses (SA)~\cite{sa1, sa2,  bpa, analysis1,sa3 }, such as type analysis or value-set analysis (VSA)~\cite{vsa}, achieve high recall by conservatively over-approximating call targets. This ensures soundness, but often at the cost of precision~\cite{bpa}. While this is suitable for CFI, it yields excessive false positives that impede precision-sensitive tasks such as reachability analysis and symbolic execution. In contrast, ML-based methods focus on improving precision by leveraging intra-procedural slicing and pattern matching to associate callsites with their likely callees~\cite{callee}. These approaches, however, frequently assume that direct and indirect calls share the same input feature space---spanning function signatures, code patterns, and memory interactions---overlooking the fact that they may exhibit fundamentally different feature distributions. For example, direct calls often resolve to fixed targets, making their backward or contextual features shallow and structurally stable, whereas indirect calls depend on runtime-computed pointers, producing deeper and more variable contextual dependencies. As a result, models~\cite{siamese, transformer, siamese2} trained predominantly on direct calls tend to struggle when applied to the more complex semantics of indirect calls. Without explicit supervision and calibration using indirect call ground truth, these models often suffer from poor recall. While they offer a promising direction, they do not fully eliminate false positives or guarantee soundness, limiting their reliability for precision- and security-critical applications.

In practice, both approaches face challenges: static analysis floods the analyst with noise, and ML risks missing critical behavior. At the same time, mainstream RE tools like IDA~\cite{idapro}, Ghidra~\cite{ghidra}, and Angr~\cite{angr} still rely on shallow heuristics, simplistic constant propagation, or impractical dynamic emulation, all leading to inaccurate CFGs. This situation forces analysts to choose between noisy static results and uncertain ML predictions, with limited options for flexible precision-recall configurations to meet diverse security needs.

\noindent\textbf{Limitations.} We identify three primary limitations in existing approaches. First, there remains a disconnect between static analysis and ML techniques; existing work fails to combine the reliability of static reasoning with the adaptability of learning-based models. Second, ML methods often rely on oversimplified features, limiting generalization from direct to indirect calls and requiring either costly supervision or sacrificing soundness. E.g., they lack scalable inter-procedural support. Third, reverse engineering tools remain technically behind academic solutions and lack the precision–recall flexibility, leaving them ill-equipped to support diverse real-world analysis needs.


\noindent\textbf{Solution:} We introduce \sysname{}, a hybrid framework for indirect call resolution that enhances precision while minimizing recall loss and preserving the soundness benefits of static analysis. 
It begins with a value-set analysis (BPA) to conservatively approximate potential callees. This is followed by a lightweight, learning-based soft signature matcher (iScoreGen), trained on direct call pairs to learn abstract compatibility features such as argument structures and memory access patterns, which generalize well to indirect calls even in stripped or highly optimized binaries. We then apply a selective inter-procedural backward analysis (iScoreRefine) to extract essential code context and memory-code dependencies. These insights are used to readjust the initial prediction scores, reinforcing or suppressing ML outputs based on supporting static evidence. The final output, \pIndGraph, annotates indirect edges with confidence scores, enabling downstream tools to adjust precision–recall trade-offs for diverse security- and precision-critical tasks.


\noindent\textbf{Contributions:}
We present \sysname{}, a multi-layered framework that bridges static soundness and machine-learned precision for indirect call resolution. Our key contributions include:
\begin{itemize}[nosep]
\item \textbf{Hybrid Architecture:} A unified design that integrates sound static analysis with learning-based refinement.
\item \textbf{Zero-Supervision Learning:} A novel feature decoupling strategy enabling generalization from direct to indirect calls without dynamic training supervision.
\item \textbf{Inter-procedural reasoning:} A scalable backward analysis that recovers long-range inter-procedural pointer dependencies to validate ML predictions.
\item \textbf{Confidence-Annotated \& Context-Adaptable CFG:} A flexible output (\pIndGraph) supporting dynamic CFG customization across application needs.
\item \textbf{Empirical Validation:} \sysname{} reduces Average Indirect Call Targets (AICT) by 44.3\% over BPA with only 0.4\% recall loss. In recall-preserving settings, it achieves 18–48.5\% more AICT reduction than prior tools at higher recall. In F1-optimized mode, it delivers 82–97\% greater reduction while sustaining 80\% F1 and 82–89\% recall, outperforming existing systems.  Our artifacts are available at: \href{https://figshare.com/s/c68d1e5ca3a18b5c3c33}{Link}.


\end{itemize}

\section{Background and Related Work}
Despite decades of work, indirect call resolution remains a trade-off between soundness and precision. Static analysis over-approximates, ML models under-generalize and are incomplete, and RE tools rely on brittle heuristics.  None provides the flexibility to balance recall and precision across diverse applications. We review key approaches---program analysis (\bpa{}~\cite{bpa}, \schedex{}~\cite{SchedExec}, BinDSA~\cite{bindsa}), learning-based (\callee{}~\cite{callee}, \attncall{}~\cite{Attncall}), and RE tools---highlighting their techniques and limitations.

\textbf{Analysis based approaches:} Analysis-driven techniques remain foundational for indirect call resolution, valued for their soundness and explainability---crucial for security tasks like CFI. When carefully engineered, they produce sound but coarse call graphs, sacrificing precision and adaptability. We examine three representative systems: \schedex{}~\cite{SchedExec}, \bpa{}~\cite{bpa}, and BinDSA~\cite{bindsa}, each exposing design trade-offs that motivate more flexible solutions. 

\schedex{}~\cite{SchedExec} combines dynamic execution with static IR-level replay, offering a pragmatic hybrid approach. However, it suffers from premature pruning and simplistic assumptions—e.g., assuming function pointers residing in contiguous memory entries. This fails in real binaries with sparse arrays, nested structures, or obfuscated layouts. \schedex{} ignores aliasing, stack arguments, and inter-procedural flows, and cannot handle recursive memory indirection. Most critically, \schedex{} performs pruning before a full value-set analysis (VSA), removing program paths without a conservative control-flow target approximation. This undermines soundness: once potential targets are prematurely removed, no post-hoc analysis can recover them. Its inability to handle complex benchmarks like \texttt{gcc} further limits its applicability.

BinDSA~\cite{DSA} adapts Data Structure Analysis (DSA)~\cite{dsa_f} to infer structures and points-to sets from lifted IR. During points-to analysis, it prunes indirect call candidates at the callsite using the recovered structure information. Even setting aside IR-lifting noise, its premature pruning leads to significant recall drops in large benchmarks such as gcc. Moreover, errors in data-structure recovery cause many valid callsites to remain unresolved.

\bpa{}~\cite{bpa} uses a fully static, Datalog-based pipeline~\cite{souffle} to build scalable and sound CFGs. Its key abstraction partitions memory into disjoint blocks, assuming pointer arithmetic stays within block bounds. This avoids reasoning about intra-block offsets and enables efficient, conservative points-to analysis suitable for large stripped binaries and CFI. However, block-level modeling leads to severe over-approximation and relies on fragile heuristics for memory partitioning. While DISA~\cite{disa} improves this with ML-driven block recovery, the remaining precision gap still causes non-trivial over-approximation for indirect call resolution. 


\textbf{ML-driven approaches:} Recent ML-based approaches like CALLEE~\cite{callee} and AttnCall~\cite{Attncall} aim to overcome static analysis over-approximation by learning callsite–callee compatibility through intra-procedural, register-level slicing. While they offer automation and fast inference, their reliance on slicing poses major limitations, especially in real-world binaries.  Small callees often lack enough instructions to generate meaningful slices; CALLEE drops these entirely, missing valid targets. Both struggle with indirect calls to short wrappers or optimized code—common in performance-critical binaries. Their intra-procedural view ignores inter-procedural flows, aliasing~\cite{aliaspointsto}, and code–memory interactions, limiting their ability to capture long-range dependencies. Their lack of explainability further reduces suitability for security-sensitive tasks~\cite{cfi1, cfi2}.

Both methods target x86-64~\cite{x861, x862} and follow similar pipelines—extracting context slices from direct call pairs and predicting indirect links. CALLEE uses a Siamese network with limited indirect fine-tuning; AttnCall employs a transformer~\cite{transformer} trained solely on direct calls. Both rely on the flawed assumption that intra-procedural, register-level program slices for direct and indirect calls share similar context patterns. 
In practice, unlike direct calls, the backward context of indirect calls often involves memory-resident function pointers, aliased stack frames, or control/data dependencies across procedures—patterns often poorly captured by local slicing. AttnCall suffers from steep recall drops (as low as 70\%) when handling indirect calls without fine-tuning. While CALLEE introduces fine-tuning, it requires costly dynamic ground truth and still fails when slices are missing or pointers are complex. AttnCall inherits these weaknesses while worsening generalization by omitting the fine-tuning phase. Its deeper model fails to compensate for missing structural reasoning. As a result, these models are less suited for high-precision or security-sensitive applications.

\textbf{RE Tool’s approaches:} Mainstream reverse engineering tools like IDA~\cite{idapro}, Ghidra~\cite{ghidra}, and Angr~\cite{angr} resolve indirect calls using heuristics, limited constant propagation, or costly emulation, resulting in under-approximated target sets and incomplete CFGs. Among them, Angr performs best due to its emulation and value-set analysis backend.
However, Angr lacks inter-procedural reasoning in the analysis and suffers from high overheads during emulation, leading to slowdowns even on small binaries. These limitations make it impractical for large or stripped binaries that require scalable, analysis-aware CFG construction with controllable soundness–precision trade-offs.



\section{Motivation and Overview} 

Resolving indirect calls remains challenging as no existing approaches simultaneously achieve soundness, precision, and scalability.  
By examining their limitations, we extract a series of design insights that underpin \sysname{}.

\indent\textbf{Neither SA nor ML Suffices Alone – A Unified Approach is Essential:}\textit{ Only a tightly integrated SA–ML framework can balance recall, precision, and scalability for indirect call resolution. Neither approach alone can achieve the balanced accuracy required for real-world analysis.}
To bridge this gap, \sysname{} employs a multi-layered architecture that lets SA do what it does best---ensure sound over-approximation---and lets machine learning do what it is good at---learn patterns that guide pruning and refinement.

\indent\textbf{Post-VSA Pruning for Soundness:} \textit{Conservative approximation must be preserved first, with pruning applied only after a sound over-approximation has been constructed.} To bridge this gap, \sysname{} adopts BPA’s block-based VSA as Layer 1 (\layervsa{}), ensuring soundness by collecting a superset of indirect callees. However, traditional VSA can be expensive and imprecise, especially in large or stripped binaries. BPA’s block abstraction offers practicality and high recall but generates more than 40\% false positives due to coarse modeling. To achieve actionable precision while retaining coverage, Layer 2 (iScoreTune) introduces a learning-based pruning strategy, while keeping Layer 1 modular enough to be replaced by any sound CFG generator.

\indent\textbf{Generalization Requires More Than Slicing:} \textit{Achieving generalization across fundamentally different call structures requires more than shallow slicing. Generalization requires separating stable signature features from indirect-specific, context-rich behaviors. Without integrating memory reasoning, contextual validation, and inter-procedural flows, ML-based solutions remain brittle.} To bridge this gap, \sysname{} adopts a feature separation strategy that distinguishes between generalizable and context-sensitive patterns essential for accurate indirect call resolution. It leverages function-signature features---such as argument count and register usage, and more---that remain consistent across direct and indirect calls, enabling robust generalization. In parallel, it extracts context-sensitive features from backward slices, such as pointer dereferences and memory loads, which capture the nuanced behavior of indirect calls. This separation forms the basis of Layer 2 (iScoreTune), which includes (L2a) iScoreGen, a learning-based module for soft signature scoring, and (L2b) iScoreRefine, a targeted inter-procedural analysis that adjusts scores based on code and memory dependencies.

\indent\textbf{Layer 2a: iScoreGen – The Role of Soft Signatures:} \textit{Robust generalization begins with soft signature matching~\cite{analysis4, softsignature}, which abstracts away structural noise and compiler-induced variations that static analysis cannot handle, making ML the only viable approach.} Matching function signatures across callsites is challenging, especially in stripped binaries where optimizations break conventions. iScoreGen addresses this issue through soft signature matching. It employs a lightweight machine learning model trained on pairs of direct calls to learn compatibility patterns that remain consistent across both direct and indirect calls. This setup operates without the need for time-consuming, exhaustive, indirect call ground truth, producing probabilistic scores that are resilient to structural variations. 
However, signature-based reasoning alone is ambiguous, as both true and false callees may share the same prototypes. Therefore, \layerScoreRefine{} validates or adjusts ML predictions using static backward memory and control-flow context.

\indent\textbf{Layer 2b: iScoreRefine – Scalable Inter-Procedural Reasoning:} \textit{A practical inter-procedural solution must balance depth and scalability by selectively applying inter-procedural refinement where it matters most.} Fully inter-procedural backward analysis is prohibitively expensive, while existing ML tools lack context awareness. iScoreRefine applies selective inter-procedural analysis only when callsite arguments align with the caller. It inspects code and memory along these paths to capture memory–control-flow coupling, readjusting ML scores, boosting confidence when dependencies match, or penalizing otherwise. This selective strategy forms a tightly coupled ML–SA framework that achieves the precision and scalability needed in the real world.

\indent\textbf{Different Needs of Downstream Applications:} \textit{ Fully accurate indirect call resolution is rarely achievable in practice. Until exact resolution is achievable, indirect call resolution during CFG construction needs to be flexible, allowing downstream analysis to adjust confidence levels based on precision or recall requirements.} 
Both academic and RE tools treat CFG edges as binary, failing to satisfy diverse precision–recall needs. For example, symbolic execution benefits from pruned CFGs, while CFI or malware analysis demands higher recall, even at the cost of false positive edges. p-IndirectCFG addresses this by assigning confidence scores to indirect edges (derived from iScoreGen and iScoreRefine). Downstream tasks can focus on high-confidence edges for precision or include lower-confidence edges when exhaustive coverage is needed.

\begin{table*}[t!]
\centering
\scriptsize
\caption{Design Comparison of ML-Based Tools vs \sysname{}}
\label{tab:academic_tool_comparison}
\begin{tabular}{|p{4.2cm}|p{3.2cm}|p{3.2cm}|p{5.2cm}|}
\hline
\textbf{Aspect} & \textbf{CALLEE}~\cite{callee} & \textbf{AttnCall}~\cite{Attncall} & \textbf{\sysname{}} \\
\hline
\textbf{Indirect call Related Ground Truth Needed During Training} & Yes & No & No \\
\hline
\textbf{Feature Design} & Mixed arg/context slices & Mixed arg/context slices & Segregated: signature (L2a) + context (L2b) \\
\hline
\textbf{Model Type} & Siamese + fine-tuning & Transformer (large, opaque) & Lightweight 4 hidden-layer DNN; no fine-tuning; fast inference \\
\hline
\textbf{Direct-to-Indirect Generalization} & Weak; direct $\neq$ indirect & Claimed generalization; empirically weak & Strong: trained on direct, refined statically \\
\hline
\textbf{Inter-procedural Support} & Limited (callee--caller only) & None & Selective backward + memory sweep \\
\hline
\textbf{Small function Support} & No & No & Yes \\
\hline
\end{tabular}
\end{table*}

\begin{figure*}[t!]
    \centering
    \includegraphics[width=0.75\linewidth]{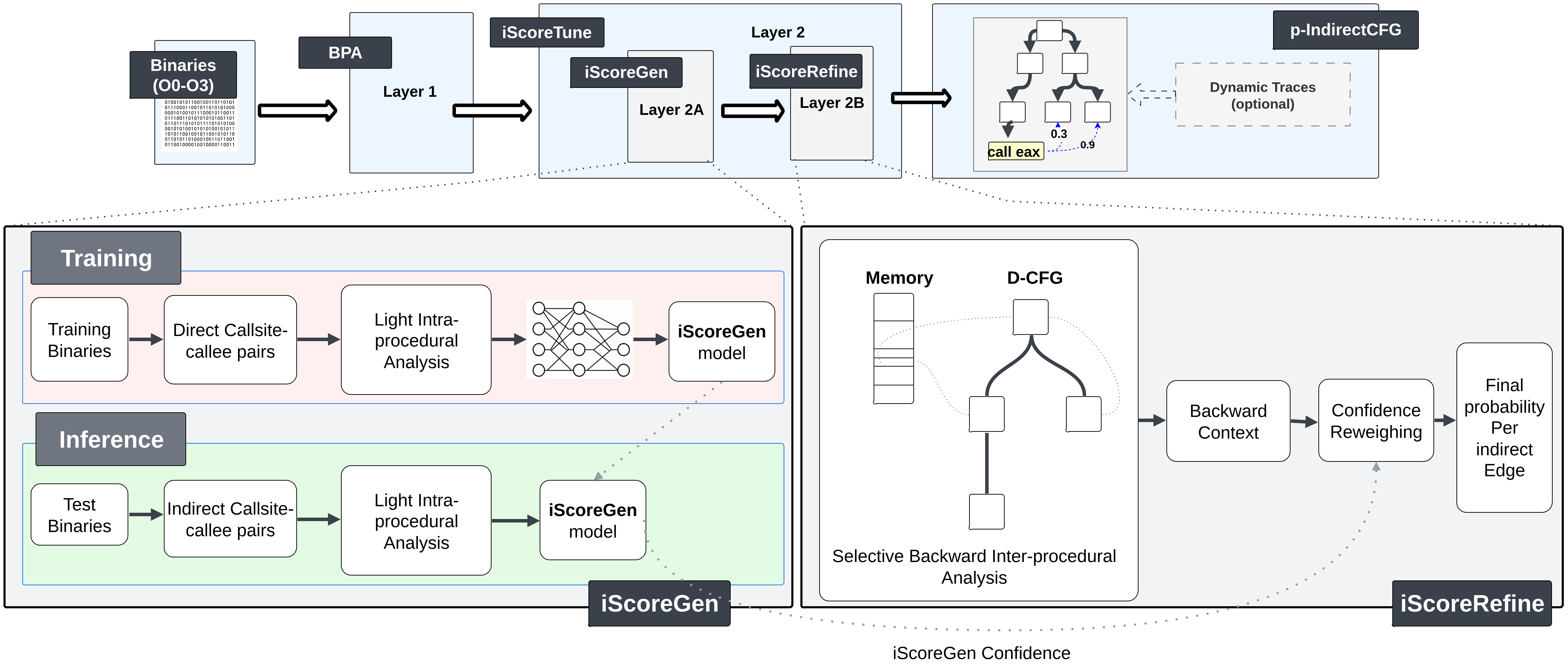}
    \caption{System Overview of the \sysname{}}
    \label{fig:overview}
\end{figure*}

Before introducing the full workflow of \sysname{}, we outline how our design differs from prior ML-based approaches (Table~\ref{tab:academic_tool_comparison}). Unlike CALLEE and AttnCall, which assume that callsite and callee slices share a similar feature space, \sysname{} separates the space into (1) a function-signature space and (2) a backward-context space. Function-signature features are stable across direct and indirect calls, making a soft signature scorer trained with direct callsite-callee pairs applicable for indirect calls. In contrast, backward-context features are often distinct between direct and indirect calls; patterns learned from direct calls may not apply. Thus, we choose to statically analyze the backward context to reduce the false positives caused by the scorer. Since critical context often lies along inter-procedural paths, \sysname{} performs inter-procedural backward traversal for indirect calls, while being selectively to improve scalability. This combined feature design improves generalization, avoids unnecessary analysis for direct calls, and enables a lightweight DNN, which is not achieved by prior approaches.

\indent\textbf{Workflow:} Figure~\ref{fig:overview} presents the workflow of \sysname{}. Given a stripped binary, \sysname{} proceeds through three stages: \textbf{Layer 1 (\layervsa{}):} A BPA-inspired block-level VSA computes a sound superset of callees.  
\textbf{Layer 2 (\layerScoreTune{}):}  
\textit{L2a iScoreGen:} For each callsite–callee pair, features are extracted and scored by a lightweight ML model.  
\textit{L2b iScoreRefine:} For all those pairs, selective backward slicing validates ML predictions while avoiding costly full analysis.  
The final CFG, called \textbf{p-IndirectCFG}, annotates indirect edges with confidence scores, enabling precision--recall trade-offs. A threshold $\in [0,1]$ controls pruning: a \emph{lower} threshold yields a more sound CFG, and a \emph{higher} threshold yields a more precise CFG. In addition, an optional dynamic-analysis module provides dynamic traces to quantify recall retention across thresholds for a given binary.

\section{Methodology}
\subsection{Layer1: Design of \layervsa{}}
We begin our pipeline with BPA~\cite{bpa}, a block-level points-to analysis framework, to conservatively over-approximate indirect call targets for soundness. This process involves creating a coarse-grained memory block model from a direct control-flow graph and performing iterative value tracking and reachability propagation until convergence. The resulting fixpoint yields a conservative set of candidate targets for each indirect callsite. Furthermore, by using reachability-based filtering, it effectively eliminates indirect calls in dead code, a feature not covered by existing ML frameworks. For implementation details, we refer to the original BPA framework.

\subsection{Layer2: Design of \layerScoreTune{} }
To refine \layervsa{} candidates, \layerScoreTune{} (i) scores callsite–callee pairs via ML-based soft signature matching (\layerScoreGen{}), and (ii) validates pairs through inter-procedural analysis (\layerScoreRefine{}), effectively separating reusable signature patterns from context.

\subsubsection{Layer2A: Design of \layerScoreGen{}}


iScoreGen evaluates callsite–callee compatibility using intra-procedural data-flow features that capture key behavioral patterns. The callsite feature extraction algorithm is in Algorithm~\ref{alg:callsite-feature}. First, \textsc{TrackIndirectCall} locates the indirect callsite in the binary and stores nearby instructions that are used later to understand the argument usage pattern. 
\textsc{LiveVariableAnalysis} identifies variables (such as registers, globals, or stack slots) that become dead (i.e., their values are not used) after the callsite, and these are treated as possible implicit arguments. \textsc{ExtractArgumentsFromLiveVars} analyzes the live variable set to identify which ones serve as function arguments and derives their feature sets. For each identified argument, \textsc{ClassifySource} determines its origin (e.g., stack, register, or global variable), which provides the context on how the value is created and passed. \textsc{AnalyzeMemoryOperand} checks whether the argument is used through address-taking or dereferencing operations, which are used to determine if it is a pointer or not. \textsc{DetectPrePostChecks} examines the surrounding instructions to see if the argument is validated by comparisons, bounds checks, or null checks before or after the callsite. \textsc{InferProbableType} assigns a coarse type hint (e.g., integer, char *, etc.) based on usage patterns and representation, complementing the structural features with semantic information. 
Finally, \textsc{CheckReturnUsage} verifies the existence of the return value of the call.  

For example, consider a minimal program shown in Listing~\ref{lst:callsite-example}, including the assembly code and its corresponding source code in comments. 
In \texttt{main}, a function pointer \texttt{fp} is set to \texttt{target\_func} (line 2). Local variables include \texttt{a = 42} (argument), \texttt{b = 100}, and a stack buffer \texttt{buf[]} (argument). The indirect call at line 12 \texttt{fp(a, buf)} (only called if $\texttt{a} > 0$) stores its return in \texttt{r}, prints it (line 15), and later prints \texttt{b} (line 16).  
Variables that are dead after the indirect callsite are highlighted in yellow, and the ones that are still being used after the callsite are in green, while, for brevity, only the most relevant variables are marked.
Here, the stack slot \texttt{-0x20(\%ebp)} (\texttt{a}) and buffer \texttt{-0x48(\%ebp)} (\texttt{buf[]}) are dead and passed as arguments via \texttt{push -0x20(\%ebp)} and \texttt{push \%eax} (address of \texttt{buf}).  

For each argument, features are then extracted (lines~5--8 in Algorithm~\ref{alg:callsite-feature}). For variable \texttt{a} in the example, \textsc{ClassifySource} marks it as a stack local. It is written with an immediate constant, used in a comparison (\texttt{cmpl \$0x0, -0x20(\%ebp)}), and passed by value, indicating a non-pointer scalar (\textsc{AnalyzeMemoryOperand}); the coarse type hint is inferred as integer (\textsc{InferProbableType}). For \texttt{buf}, the source is also the stack, but its address is taken (\texttt{lea -0x48(\%ebp), \%eax; push \%eax}), so pointer-likeness is true. No validation checks are observed in \textsc{DetectPrePostChecks}, unlike the other argument, and as a stack-allocated character array initialized with a string literal, its coarse type hint is a character pointer. Return usage (\textsc{CheckReturnUsage}, line~10 in Algorithm~\ref{alg:callsite-feature}) shows the result in \texttt{\%eax} is saved at \texttt{-0x28(\%ebp)} and immediately printed, highlighted in purple. Finally, the analysis recovers two arguments, each annotated with origin, pointer-likeness, validation, and type hints. Together with the return usage, they form the callsite’s feature vector (\textsc{AggregateCallsiteFeatures}, line~11 in Algorithm~\ref{alg:callsite-feature}).

\SetKwComment{Comment}{$\triangleright$\ }{}
\SetCommentSty{scriptsize\color{blue}}

\begin{algorithm}[t!]
\caption{Callsite Feature Extraction (iScoreGen)}
\scriptsize
\label{alg:callsite-feature}
\KwIn{Binary $B$, Indirect Callsite $c$}
\KwOut{Feature vector $F_c$ representing $c$}

$R \leftarrow$ \textsc{TrackIndirectCall}($B$, $c$) \algcomment{Track related info of the indirect call}

$L \leftarrow$ \textsc{LiveVariableAnalysis}($B$, $c$, $R$) \algcomment{Find variables dead after the callsite R}

$A \leftarrow$ \textsc{ExtractArgumentsFromLiveVars}($L$, $R$) \algcomment{Infer arguments from dead variables at $R$.}

\ForEach{$a \in A$}{
    \Indp
    $a.\text{origin} \leftarrow$ \textsc{ClassifySource}($a$)  \algcomment{Determine source: stack, global, register, more}
    
    $a.\text{pointer\_likeness} \leftarrow$ \textsc{AnalyzeMemoryOperand}($a$) \algcomment{infer pointer behavior}
    
    $a.\text{validation\_info} \leftarrow$ \textsc{DetectPrePostChecks}($B$, $a$, $c$)  \algcomment{Check if the argument is validated near the call}
    
    $a.\text{type\_hint} \leftarrow$ \textsc{InferProbableType}($a$)  \algcomment{Infer a coarse type based on usage and representation}
    
    \Indm
}

$arg\_count \leftarrow |A|$ \algcomment{Number of inferred arguments}

$ret\_used \leftarrow$ \textsc{CheckReturnUsage}($B$, $c$) \algcomment{Detect if return value is used}

$F_c \leftarrow$ \textsc{AggregateCallsiteFeatures}($A$, $arg\_count$, $R$, $ret\_used$)

\Return{$F_c$}
\end{algorithm}


\begin{lstlisting}[float, style=asmgas,linewidth=\columnwidth,caption={Example of Callsite Feature Extarction}, label={lst:callsite-example}]
    080491be <main>:
    080491d6:  movl   $0x8049196, (*@\colorbox{yellow}{-0x1c(\%ebp)}@*); int (*fp)(int, char *) = target_func;
    080491dd:  movl   $0x2a,      (*@\colorbox{yellow}{-0x20(\%ebp)}@*); int a = 42;
    080491e4:  movl   $0x64,      (*@\colorbox{green!30}{-0x24(\%ebp)}@*) ; int b = 100;
    080491eb:  lea    (*@\colorbox{yellow}{-0x48(\%ebp)}@*), %edi ; buf destination
    080491ee:  mov    $0x804a024, %esi; string literal
    080491f3:  rep movsl; char buf[] = "..."; 
    08049200:  cmpl   $0x0, (*@\colorbox{yellow}{-0x20(\%ebp)}@*)   ; if (a > 0) then fp
    08049209:  lea    (*@\colorbox{yellow}{-0x48(\%ebp)}@*), %eax ; prepare buf address
    0804920c:  push   (*@\colorbox{yellow}{\%eax}@*) ; arg2 buf
    0804920d:  push   (*@\colorbox{yellow}{-0x20(\%ebp)}@*) ; arg1 a
    08049213:  call   *-0x1c(\%ebp) ; r = fp(a, buf);
    08049218:  mov    %eax, (*@\colorbox{purple!50}{-0x28(\%ebp)}@*); save return r
    0804921e:  push   (*@\colorbox{purple!50}{-0x28(\%ebp)}@*)
    08049226:  call   printf@plt ; printf("%d", r)
    08049231:  push   (*@\colorbox{green!30}{-0x24(\%ebp)}@*) ;start of print("%d\n", b)
\end{lstlisting}

On the callee side (Algorithm~\ref{alg:callee-feature}), we identify likely parameters of the function by detecting variables using \textsc{GetAllStackAndRegisterVars} and mark variables read before any write using \textsc{IsUsedBeforeWritten}, indicating values expected from the caller. In stripped or highly optimized binaries, this task is challenging due to effects such as inlining and register reuse, and we observe that traditional tools (e.g., Angr) often fail to reliably identify such arguments. To resolve this and account for cases where some callsite arguments are unused by the callee (e.g., optional parameters), each candidate receives an additive score \textsc{ComputeArgumentHeuristicScore} using cues like positive stack offset, early use, never-written, cross-basic-block use, high access frequency, and direct memory access at a positive offset (More details in Appendix in Table~\ref{tab:heuristics}); candidates with cumulative score $\ge \tau$ are retained as likely arguments (we set $\tau=6.0$ in our experiments). In the running example \texttt{target\_func} (Listing~\ref{lst:callee-example}), we highlight the variables used before being written. 
In our example, \texttt{a} at \texttt{0x8(\%ebp)} scores $2.0$ (stack positive offset access) $+$ $1.5$ (as never written) $+$ $2.0$ (as used early in the function) $+$ $0.0$ (single basic block use only) $+$ $0.0$ (as used less than thrice) $+$ $2.0$ (as direct mem at $+\,$offset, this is included for pointer indirection) $= 7.5$. \texttt{buf} at \texttt{0xC(\%ebp)} scores $2.0$ (positive offset) $+$ $1.5$ (never written) $+$ $2.0$ (early use) $+$ $0.0$ (single basic block) $+$ $0.0$ (frequency not high) $+$ $2.0$ (direct mem at $+\,$offset) $= 7.5$. Return detection uses an analogous rule on \texttt{\%eax} (marked as return value if total score $\ge 2.0$): $1.0$ (as written to \texttt{\%eax} at line~8 in Listing~\ref{lst:callee-example}) $+$ $1.5$ ( as written in final basic block) $+$ $0.0$ (as no constant moved to \texttt{\%eax} towards the end of the function) $= 2.5$, so a return is present. Selected arguments then receive the same features as on the callsite---\textsc{ClassifySource}, \textsc{AnalyzeMemoryOperand}, and \textsc{InferProbableType} --- yielding the callee feature vector for learning-based compatibility scoring.

\begin{algorithm}[t!]
\small
\caption{Callee Feature Extraction (iScoreGen)}
\label{alg:callee-feature}
\scriptsize
\KwIn{Binary $B$, Function $f$}
\KwOut{Feature vector $F_f$ representing $f$}

$V \leftarrow$ \textsc{GetAllStackAndRegisterVars}($f$) \algcomment{Collect stack/register variables for inspection}

$A \leftarrow$ [\ ] \algcomment{Initialize list of inferred arguments}


\ForEach{$v \in V$}{
    \Indp
    \If{\textsc{IsUsedBeforeWritten}($v$)}{
        $score \leftarrow$ \textsc{ComputeArgumentHeuristicScore}($v$)  \algcomment{Score each probable candidate to select high probable ones}
        
        \If{$score > \tau$}{
            $a.\text{origin} \leftarrow$ \textsc{ClassifySource}($v$)  \algcomment{Determine source: stack, global, register, more}
    
            $a.\text{pointer\_likeness} \leftarrow$ \textsc{AnalyzeMemoryOperand}($v$) \algcomment{infer pointer behavior}
            
            $a.\text{type\_hint} \leftarrow$ \textsc{InferProbableType}($v$) \algcomment{Infer a coarse type based on usage and representation}
            
            $A$.append($a$) 
        }
    }
    \Indm
}

$arg\_count \leftarrow |A|$ \algcomment{Number of inferred arguments}

$ret\_present \leftarrow$ \textsc{CheckReturnUsage}($B$, $f$) \algcomment{Detect if return value is used}

$F_f \leftarrow$ \textsc{AggregateCalleeFeatures}($A$, $arg\_count$, $ret\_present$)

\Return{$F_f$}
\end{algorithm}

\begin{lstlisting}[float, style=asmgas,linewidth=\columnwidth, escapeinside={(*@}{@*)},
caption={Example of Callee Feature Extraction}, label={lst:callee-example}]
    08049196 <target_func>:
    80491a3: push   (*@\ubw{0xc(\%ebp)}@*) ; buf (use)
    80491a6: push   (*@\ubw{0x8(\%ebp)}@*) ; a   (use)
    80491a9: push   $0x804a008 ; format string
    80491ae: call   8049070 <printf@plt>
    80491b3: add    $0x10,%esp
    80491b6: mov    0x8(%ebp),%eax ; use a -> eax
    80491b9: add    $0xa,%eax ; a + 10
    80491bd: ret ; return a+10
\end{lstlisting}

The training phase of iScoreGen uses labeled callsite–callee pairs obtained from direct calls in training binaries. For each pair, we extract features using the callsite and callee feature extraction algorithms and combine them into a single feature vector. This vector is fed into a lightweight DNN that models soft compatibility between callsites and callees, outputting a sigmoid-based probability score. During inference, Layer 1’s BPA generates candidate indirect callsite–callee pairs. iScoreGen computes their features and feeds them into the trained DNN to produce compatibility scores, which are later refined by inter-procedural static analysis in iScoreRefine.

\subsubsection{Layer2B: Design of \layerScoreRefine{}} \label{sec:layer2b}
To strengthen prediction beyond iScoreGen, we employ a selective inter-procedural backward analysis. This step recovers long-range pointer arithmetic and recursive memory chains that influence indirect call operands—dependencies often missed by intra-procedural or ML-based approaches. These recovered traces provide static evidence for confirming or discarding potential callsite–callee pairs.

The process, outlined in Algorithm~\ref{alg:backward-analysis}, computes possible targets for the operand of each indirect callsite. It begins by locating the function and basic block containing the callsite. Using this location as a seed, we perform a bounded inter-procedural backward traversal (\textsc{ExtractBackwardPaths}, line~6) over the direct control-flow graph (DCFG), with the bound expressed as a height limit on the number of basic blocks explored backwards. Along each backward path (lines~7–16), we examine instruction operands: if an operand directly corresponds to a known function start address, we classify it as a control-flow-resolved function pointer. If the operand refers to global memory (e.g., in \texttt{.data}, \texttt{.rodata}, or \texttt{.bss}), we invoke \textsc{RecursiveMemorySweep} to traverse the memory region. This resolution follows pointer chains up to a configurable recursion depth, thereby recovering function entry points that are only indirectly referenced. Instead of relying on BPA’s heuristic memory blocks, we leverage DISA’s~\cite{disa} ML–driven flexible memory layout, which generalizes better across binaries.

To expand the search context, we incorporate recursive cross-reference reasoning via \textsc{CrossRefResolve}. Specifically, we identify callsites that invoke the current function containing the indirect call and examine whether those sites push meaningful operands (e.g., pointers or addresses) as arguments. When such evidence exists, we apply a one-basic-block backward analysis at the cross-referenced caller and repeat the inspection steps (lines~7–16) as in the original callsite traversal. We then recursively analyze upstream cross-referenced callsites using the same backward traversal and memory sweep strategy, bounded by a maximum recursion depth. This enables the analysis to capture symbolic pointer flows originating outside the immediate function scope.

\begin{figure}
    \includegraphics[width=1\linewidth]{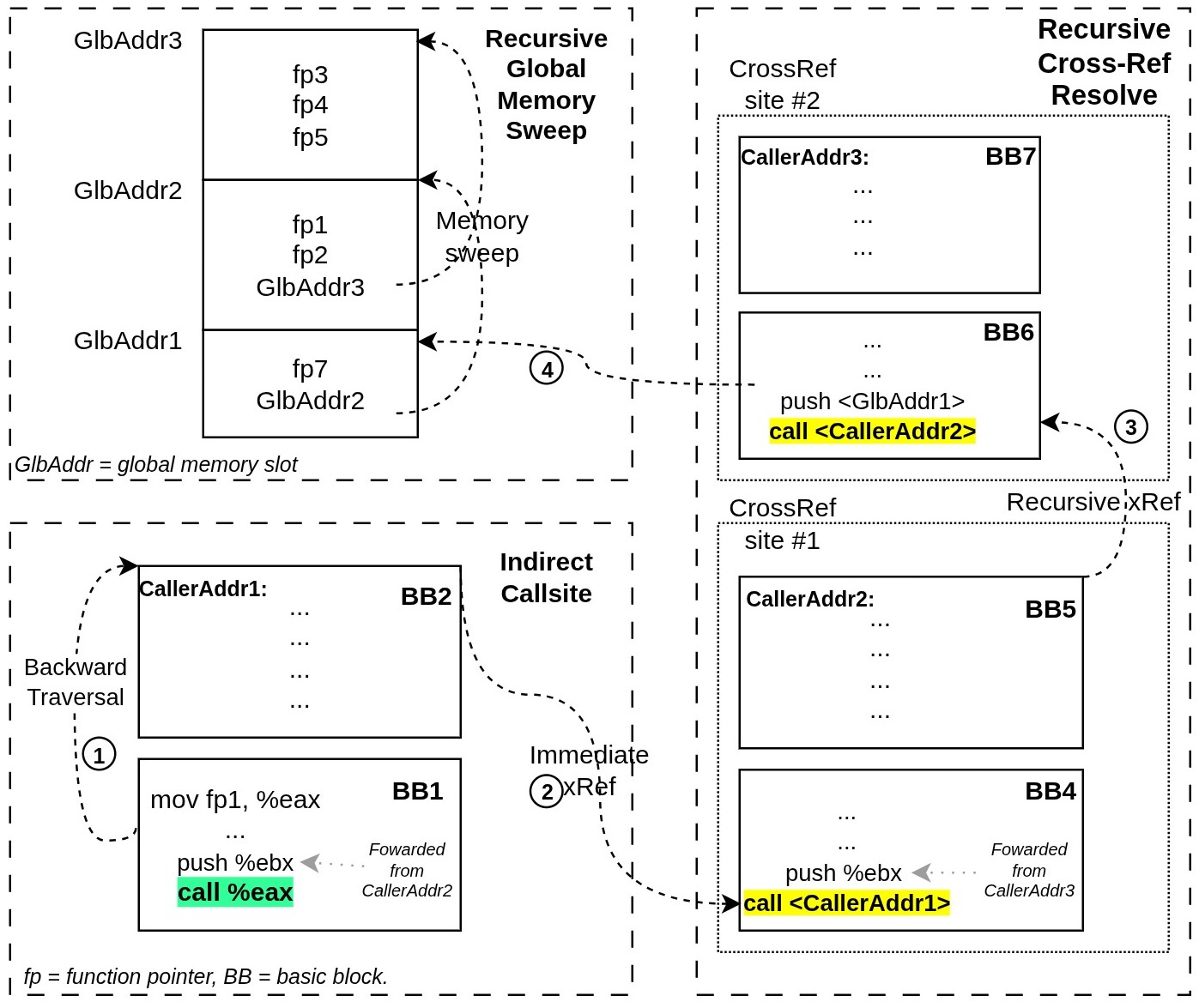}
    \caption{Example of L2b analysis using backward traversal, cross-reference resolution, and global memory sweep.}

    \label{fig:back_Algo_Example}
\end{figure}

In Figure~\ref{fig:back_Algo_Example} we illustrate via an example where the indirect callsite begins at BB1.  The argument pushed at this callsite, however, originates from a long inter-procedural chain: it is first introduced at CallerAddr3, propagated through CallerAddr2, forwarded again at CallerAddr1, and finally consumed at the indirect callsite.  First, we extract the backward path starting with BB1 (edge~\circled{1}) and analyze its instructions. During this process, we encounter a move instruction loading a function pointer (fp1) into \%eax; we mark this as a control-flow-resolved pointer and store it in the callsite’s candidate map. Following cross-reference resolution (edges~\circled{2}-\circled{3}), we detect that CallerAddr1 itself is invoked by CallerAddr2 (BB4), which passes along pointer operands, and further upstream by CallerAddr3 (BB6), which pushes GlbAddr1 as an argument. At this point, we apply the DISA-inspired global block model to recursively sweep the global memory slot starting at GlbAddr1 and discover fp7 as a valid function pointer (edge~\circled{4}). The sweep also reveals a pointer chain to GlbAddr2, which is then recursively explored until no further chains are found or the recursion depth limit is reached. The same recursive strategy applies to cross-reference resolution as well. Finally, to maintain practicality for large binaries, if the analysis time of any callsite exceeds a predefined bound, we retain the partial results and proceed to the next callsite.

Our analysis is designed to be both adaptable and time-bounded. Parameters such as control-flow traversal depth, recursive memory sweep levels, and cross-reference exploration can be adjusted to balance precision, coverage, and runtime. 
Deeper settings yield broader analysis and better coverage but require more time, creating a trade-off between  efficiency and completeness. Each callsite is subject to a hard timeout to prevent bottlenecks when analyzing large binaries.

After identifying relevant function pointers for a callsite, we refine iScoreGen’s prediction scores: if the callee appears in the recovered set, its score is increased by a score (e.g., +0.1); if absent, it is decreased by the same amount and clipped to [0,1]. This modest adjustment ensures that static evidence acts as a soft correction, complementing rather than overriding the ML model’s confidence distribution. For stricter separation between true and false targets, larger adjustments (e.g., 0.5) allow more aggressive pruning at a modest cost to recall. These mechanisms give users the flexibility to balance precision and performance according to their analysis needs.

\begin{algorithm}[t!]
\scriptsize
\caption{Selective Backward Inter-Procedural Analysis (\layerScoreRefine{})}
\label{alg:backward-analysis}
\KwIn{ Indirect Callsite list $\mathcal{C}$; DCFG $\mathcal{G}_{CFG}$; DISA-inspired global memory blocks $\mathcal{M}$;
Function start set $\mathcal{S}$; Function ranges $\mathcal{F}$; xRef map $\mathcal{X}$;
Max CFG backward traversal height $H$; Max memory sweep depth $R$;
Max cross-reference depth $D$; Timeout $T$
}
\KwOut{Map $\mathcal{T}$ from callsite $\rightarrow$ resolved target set}

$\mathcal{T} \leftarrow \emptyset$ \algcomment{Initialize result map for all callsites} \\

\ForEach{callsite $C \in \mathcal{C}$}{
  $T_C \leftarrow \emptyset$ \algcomment{Resolved targets for callsite $C$} \\
  start\_time $\leftarrow$ \textsc{GetCurrentTime}() \algcomment{Start timer for timeout check} \\
  $f_C, b_C \leftarrow$ function and block containing $C$  \algcomment{Locate the containing function and block} \\
  $\mathcal{P}_C \leftarrow$ \textsc{ExtractBackwardPaths}($b_C$, $H$) \algcomment{Extract backward control-flow paths of basic block height H} \\

  \ForEach{path $\pi \in \mathcal{P}_C$}{
    \ForEach{instruction $i \in \pi$}{
      \If{$\textsc{Elapsed}(start\_time) > T$}{ \textbf{break all}\algcomment{Timeout exceeded for this callsite} }
      \If{operand of $i$ is immediate $\in \mathcal{S}$}{
        $T_C$.add(\texttt{fn\_ptr}) \algcomment{Record direct function pointer}
      }
      \If{operand of $i$ references memory in $\mathcal{M}$}{
        $F_{mem} \leftarrow$ \textsc{RecursiveMemorySweep}($i$, $\mathcal{M}$, $R$)\algcomment{Dereference pointer chain} \\
        \ForEach{$f \in F_{mem}$}{
          $T_C$.add(\texttt{fn\_ptr})  \algcomment{Record memory-resolved function pointer
        }
      }
    }
  }

  \textsc{CrossRefResolve}($f_C$, $T_C$, $start\_time$, $D$, 0)

  $\mathcal{T}[C] \leftarrow T_C$
}
\Return{$\mathcal{T}$}

\BlankLine
\SetKwFunction{CrossRefResolve}{CrossRefResolve}
\SetKwProg{Fn}{Function}{:}{}
\Fn{\CrossRefResolve{$f$, $T_C$, $start\_time$, $D$, $d$}}{
  \If{$d \geq D$}{ \textbf{return} }
  $\mathcal{X}_f \leftarrow \mathcal{X}[f]$  \algcomment{Get cross-referencing callsites for function $f$} \\
  \ForEach{$x \in \mathcal{X}_f$}{
    \If{arg at $x$ is pointer or memory-based}{
      \If{$\textsc{Elapsed}(start\_time) > T$}{ \textbf{break} }
      $\mathcal{P}_x \leftarrow$ \textsc{ExtractBackwardPaths}($x$, $1$) \algcomment{1-block backward slice for XREF site}  \\
      repeat analysis loop line 7-16 for each $\pi \in \mathcal{P}_x$ \algcomment{Analyze xRef paths and updates $T_C$} \\
                $f_x \leftarrow \mathcal{X}[x]$  \\
             \ForEach{$f’ \in f_x$} {
             \If{arg at $x$ is pointer or memory-based}{
            \CrossRefResolve{$f’$, $T_C$, $start\_time$, $D$, $d+1$
            }}

    }}}}
  }

\end{algorithm}

\subsection{\pIndGraph{}}
Indirect call usage varies widely across binaries, so a single score threshold cannot provide a uniform recall profile across benchmarks. This variability arises from structural and semantic factors—ranging from sparse, well-isolated dispatch sites to dense, aliased, or polymorphic patterns that cross procedure boundaries—and is further influenced by callee density, target ambiguity, compiler optimizations (e.g., inlining and merging), and mismatches between static and runtime behavior. Consequently, while the general rule holds—lower thresholds yield more sound CFGs (higher recall) and higher thresholds yield more precise CFGs (fewer spurious edges)—the optimal cutoff remains benchmark-specific.

To calibrate thresholds on a per-binary basis, we include an optional dynamic-analysis module that runs the program and logs indirect call edges as callsite–callee address pairs. Because dynamic traces are always input-dependent and incomplete, in this setting, even heavily incomplete traces are sufficient to anchor thresholds. Moreover, dynamic traces highlight the edges exercised in practice, guiding threshold calibration toward retaining critical behavior while safely pruning low-confidence edges that are unlikely to appear at runtime, ensuring the final control-flow graph remains accurate and relevant for downstream analysis. Our tool would be useful even without dynamic traces: setting the threshold to zero retains all Layer-1 candidates for soundness, whereas increasing the threshold prunes low-confidence edges to improve precision at a cost to recall.  


\section{Implementation}

\textbf{BPA (L1):}
We directly reuse the Soufflé-based~\cite{souffle} BPA implementation, which produces a sound candidate set of callees for each indirect callsite. This conservative superset forms the input for subsequent refinement. 

\textbf{iScoreGen (L2a)} leverages intra-procedural data-flow analysis using angr’s~\cite{angr} \textit{Live Variable Analysis}, \textit{Variable Recovery Fast}, and \textit{Calling Convention Analysis} to identify stack and register variables at both callsite and callee and to further identify them as potential arguments. To mitigate Angr’s imprecision at the callee side, we apply a heuristic filtering scheme (details in Appendix~\ref{appendix:heuristics}). Extracted features are encoded into fixed-length callsite–callee vectors and classified by a deep neural network (TensorFlow 2~\cite{tensorflow2015-whitepaper}/Keras~\cite{keras}) with four hidden layers, Batch Normalization, LeakyReLU activation, and Dropout. The model outputs a compatibility score via a sigmoid neuron and is trained with binary cross-entropy, adam, early stopping, and learning-rate decay (details in Appendix~\ref{appendix:ml}).

\textbf{iScoreRefine (L2b)}, conducts a bounded backward inter-procedural traversal (height = 40 Basic Blocks) on the Angr-generated CFGFast of the DCFG, combined with a recursive memory inspection (depth = 1) using angr’s cross-reference capabilities. We refine the target set by merging it with iScoreGen’s predictions, applying a conservative score adjustment of +0.1 for matches and -0.1 for mismatches to maintain recall. We also tested higher thresholds, such as 0.2 to 0.5, for more aggressive pruning.

\textbf{Dynamic Tracing Module:} We developed a PIN-based~\cite{pin} dynamic tracer to log indirect callsites and their executed callees. For SPEC2006 benchmarks, we followed the same methodology as BPA~\cite{bpa}, AttnCall~\cite{Attncall}, and BinDSA~\cite{bindsa}, using SPEC-provided input seeds to ensure consistency. Since real-world binaries (e.g., exim, lighttpd, thttpd, nginx, memcached) lack standardized I/O setups, existing tools—including CALLEE~\cite{callee}—do not provide dynamic traces. To address this, we manually designed best-effort, comprehensive input scenarios to generate representative traces. In our evaluation, to balance AICT reduction and recall retention, we selected a threshold per benchmark by analyzing the L2a and L2b output distributions over a randomly sampled 30\% of the dynamic traces. For binaries without usable inputs, our tracer integrates with AFL-Pin~\cite{aflpin}, enabling trace collection. This flexibility ensures that threshold calibration remains feasible across diverse binaries and execution contexts.

\textbf{Hyperparameters:}
 Our refinement pipeline exposes parameters that balance accuracy and runtime: CFG backward traversal height (40 Basic Blocks), memory sweep depth (1), cross-reference recursion depth (3), and a per-callsite timeout of 250 seconds. iScoreRefine adjusts scores conservatively by ±0.1 to preserve recall, though higher thresholds can be used for more aggressive pruning. The pruning threshold for p-IndirectCFG is optionally dynamic trace-driven, enabling precision–recall trade-offs---higher thresholds retain only high-confidence edges, while lower ones favor broader recall. 
 
\textbf{Dataset and Experimental Setup:} We compiled 2,664 x86-32~\cite{x32} binaries from binutils~\cite{gnu_binutils}, coreutils~\cite{gnu_coreutils}, and MiBench~\cite{mibench} using GCC 7.5. Angr was used to extract a random subset of direct callsite–callee pairs, yielding 333,063 valid and 469,872 invalid pairs, split into 80\% training and 20\% validation sets. Following prior works~\cite{bpa, Attncall, callee, bindsa}, our test dataset consists of SPEC2006 benchmarks and five real-world binaries (thttpd, memcached, lighttpd, exim, and nginx) compiled with GCC 9.2.0. On the rare occasion that Angr’s intra-procedural analysis produced no result (due to its implementation issues), we skipped the item in iScoreGen and reintroduced it downstream with its L2a score set to 0. Experiments ran on Ubuntu 22.04 (kernel 6.8.0) with 352 GB RAM and a 44-core Intel Xeon CPU.

\section{Evaluation}
We evaluate the effectiveness of \sysname{} by examining the contribution of each layer---BPA (L1), iScoreGen (L2a), and iScoreRefine (L2b)---to indirect call target resolution. Our evaluation further explores the framework’s comparative performance against academic and reverse engineering tools, and its adaptability to real-world scenarios through case studies and ablation experiments. To structure this analysis, we address five key research questions:

\begin{itemize}[nosep]
\item \textbf{RQ1:} How do iScoreGen (L2a) and iScoreRefine (L2b) complement BPA (L1) to improve indirect call target resolution?
\item \textbf{RQ2:} What is the significance of iScoreRefine (L2b)? How critical is the choice of score-adjustment threshold in L2b for balancing conservative and aggressive pruning?
\item \textbf{RQ3:} How does \sysname{} compare to state-of-the-art academic frameworks and widely used reverse engineering tools?
\item \textbf{RQ4:} What is the effect of removing individual layers or combinations of layers, as studied in our ablation analysis? 
\item \textbf{RQ5:} How does threshold sensitivity in p-IndirectCFG affect its performance trade-offs, and how does this, in turn, influence downstream analyses?

\end{itemize}

\subsection{ RQ1: Layer-wise Contribution Analysis} \label{sec:rq1}


To evaluate \sysname{}, we analyze the contributions of its three layers---BPA (L1), iScoreGen (L2a), and iScoreRefine (L2b)---using two metrics: Average Indirect Call Targets (AICT) and recall.  Following prior works~\cite{callee,Attncall, bpa, bindsa, SchedExec}, AICT serves as our precision indicator: it is the average number of predicted targets per indirect callsite (lower is better). Recall is computed against the pseudo ground truth derived from dynamic execution traces and should be read as an estimate of coverage rather than absolute recall, since unobserved-but-valid targets may exist. Our aim is to retain coverage while progressively reducing AICT; accordingly, we interpret the metrics jointly: high recall with low AICT indicates a balanced prediction set, whereas high AICT signals over-approximation.

As established in prior work~\cite{Attncall, callee}, recall is evaluated in two ways: global recall, which aggregates correct predictions across all callsites, and AICT recall, which computes and averages recall per callsite. The formal definitions of both metrics are provided in Appendix~\ref{appendix:global_recall}. Although global recall indicates overall correctness, it can be misleading when datasets include numerous easy callsites alongside a few complex ones. In such scenarios, high global recall may mask failures on critical complex cases, creating a false sense of robustness. Conversely, AICT recall treats each callsite equally, providing a more accurate representation of average performance. Therefore, we prioritize AICT recall as our main metric.

Our layer-wise evaluation (Table~\ref{tab:layerwise-results}) highlights the effectiveness of \sysname{}. We present our own re-evaluation of BPA, highlighting that L1 alone produces a conservative, over-approximated set of targets but guarantees the highest recall. Adding iScoreGen (L2a) boosts AICT precision by 19.2\% through ML-driven soft signature matching, and iScoreRefine (L2b) further improves pruning by 31\%, together delivering a 44.3\% overall improvement over BPA with almost no recall loss (0.4\%). Some AICT values fall below 1 due to reachability analysis, which filters out dead-codes, a feature not provide by prior ML works~\cite{callee, Attncall}. Benchmarks like perlbench, h264ref, gobmk, exim, and nginx show improvements of 30–75\% over static-only analysis. Compared to the BPA-hybrid design in~\cite{bpa}, which used rigid argument-count filters, \sysname{} achieves an additional 24.2\% reduction, validating its learning-based layered design. 

\subsection{RQ2: iScoreRefine --- Importance of Static Inter-Procedural Validation}

\begin{figure}[t!]
    \centering
    \includegraphics[width=\linewidth]{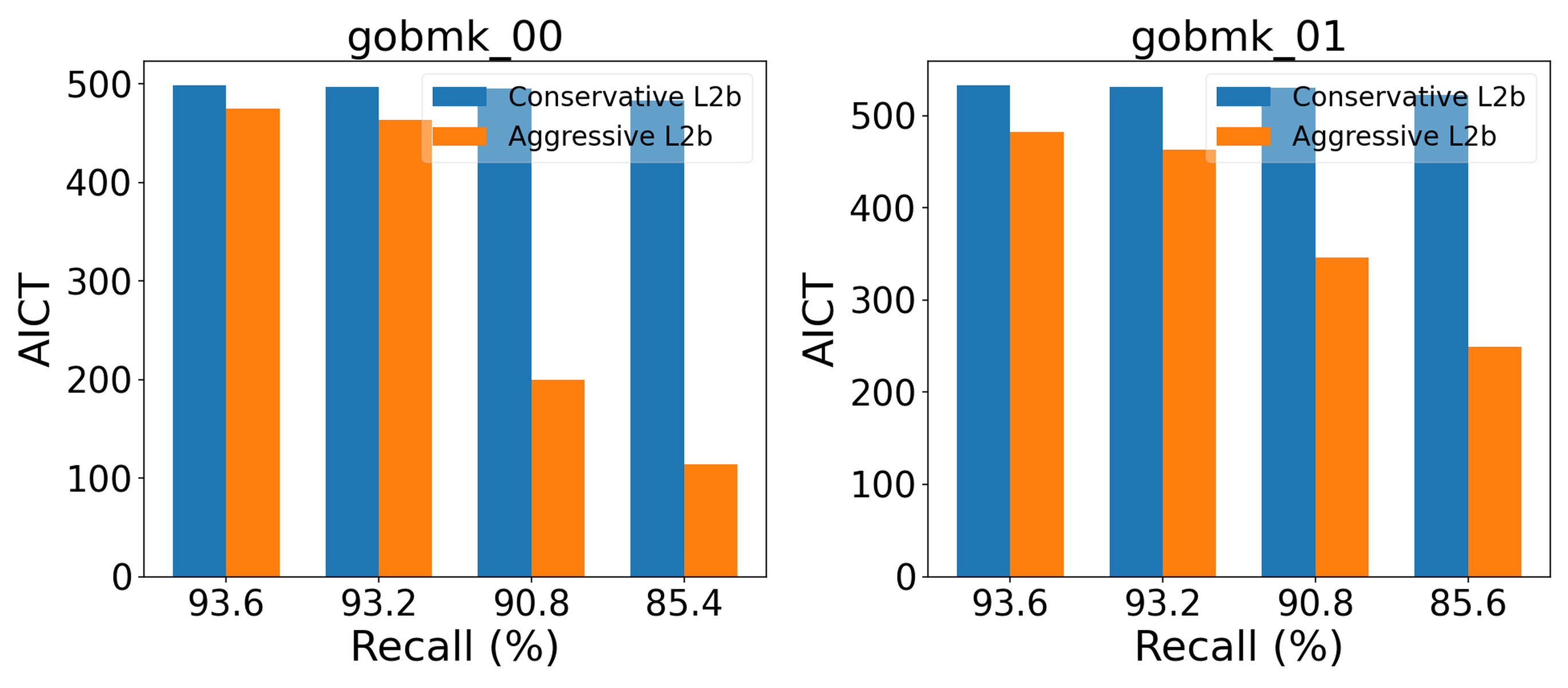}  
    \caption{Effect of L2b Pruning Strength on AICT: Aggressive vs. Conservative Across Recall Levels.}
    
    \label{fig:rq2}
\end{figure}

This section explores (i) the role of iScoreRefine (L2b) in indirect call resolution and (ii) the balance between its conservative and aggressive modes. While L2a enhances pruning with ML-driven soft signature matching, L2b employs dual strategies: conservative mode prioritizes recall, and aggressive mode accepts recall trade-offs for better precision. The advantage of L2b emerges when recall constraints are relaxed, allowing sharper pruning and significant AICT reduction—an ability lacking in existing tools. We assess L2b's effectiveness by re-scoring L2a’s outputs with varied adjustment scores. Lower score adjustment ensures lower recall loss, while higher score adjustment enables aggressive pruning, significantly reducing AICT at the cost of some recall. This approach aligns precision-recall balance with specific application needs. We illustrate this on the gobmk benchmark (Figure~\ref{fig:rq2}). We selected gobmk for its high AICT, which makes improvements visible, and present O0 and O1 for brevity. The same trend holds across benchmarks, especially in larger binaries where code complexity increases false positives. Figure~\ref{fig:rq2} shows L2b’s two modes: conservative (readjustment with $\pm$0.1) and aggressive ($\pm$0.5). As recall decreases, the gap between modes widens. In gobmk-O0, for example, at 93.6\% recall, aggressive pruning reduces AICT by 6.6\% versus conservative; at 90.8\% recall, by 59\%; and at 85\% recall, by 76\%, and so on. The effect is strongest at lower optimization levels but remains consistent across configurations. Overall, iScoreRefine enhances iScoreGen by refining predictions and offering precise control over the precision-recall trade-off.
\begin{table}[t!]
\caption{Layer-wise comparison of indirect call prediction.} 
\centering
\tiny
\begin{tabular}{l|c|c|c|c}
\hline
\textbf{Binary} & \textbf{Size} & \textbf{L1} & \textbf{L2a} & \textbf{L2b} \\
\hline
bzip2-O0 & 0.2 & 2 / 100 & 1.6 / 100 & \textbf{1 / 100} \\
bzip2-O1 & 0.2 & 2 / 100 & \textbf{1.6 / 100} & \textbf{1.6 / 100} \\
bzip2-O2 & 0.2 & 2 / 100 & \textbf{1.5 / 100} & \textbf{1.5 / 100} \\
bzip2-O3 & 0.3 & 2 / 100 & \textbf{1.5 / 100} & \textbf{1.5 / 100} \\
\hline
sjeng-O0 & 0.3 & 7 / 100 & \textbf{6 / 100} & \textbf{6 / 100} \\
sjeng-O1 & 0.4 & 7 / 100 & \textbf{6 / 100} & \textbf{6 / 100} \\
sjeng-O2 & 0.5 & 7 / 100 & \textbf{6 / 100} & \textbf{6 / 100} \\
sjeng-O3 & 0.6 & 7 / 100 & 7 / 100 & 7 / 100 \\
\hline
milc-O0  & 0.4 & 2 / 100 & 2 / 100 & 2 / 100 \\
milc-O1  & 0.5 & 2 / 100 & 2 / 100 & 2 / 100 \\
milc-O2  & 0.5 & 2 / 100 & 2 / 100 & 2 / 100 \\
milc-O3  & 0.6 & 1 / 100 & 1 / 100 & 1 / 100 \\
\hline
sphinx3-O0 & 0.5 & 1.3 / 100 & \textbf{0.8 / 100} & \textbf{0.8 / 100} \\
sphinx3-O1 & 0.7 & 1.3 / 100 & \textbf{1 / 100} & \textbf{1 / 100} \\
sphinx3-O2 & 0.7 & 0.7 / 100 & \textbf{0.6 / 100} & \textbf{0.6 / 100} \\
sphinx3-O3 & 0.8 & 0.7 / 100 & \textbf{0.6 / 100} & \textbf{0.6 / 100} \\
\hline
hmmer-O0     & 0.7 & 2.9 / 100 & \textbf{2.4 / 100} & \textbf{2.4 / 100} \\
hmmer-O1     & 1 & 4.3 / 100 & \textbf{2.8 / 100} & \textbf{2.8 / 100} \\
hmmer-O2     & 1.1 & 2.8 / 100 & \textbf{2.3 / 100} & \textbf{2.3 / 100} \\
hmmer-O3     & 1.3 & 1 / 100 & 1 / 100 & 1 / 100 \\
\hline
h264ref-O0   & 1.8 & 5.7 / 100 & 3.6 / 100 & \textbf{ 3.4 / 99.8}  \\
h264ref-O1   & 2.1 & 5.2 / 99.7  & 4.2 / 99.7  & \textbf{3.6 / 99.2}  \\
h264ref-O2   & 2.1 & 26.7 / 100 & 6.3 / 100 & \textbf{5.9 / 99.7} \\
h264ref-O3   & 3 & 18.5 / 100 & 14.7 / 100 & \textbf{6.9 / 99} \\
\hline
gobmk-O0 & 5.2 & 884.6 / 100 & 715 / 99.1 & \textbf{532.9 / 98 }\\
gobmk-O1 & 5.9 & 1336.3 / 100 & 1033 / 99.1 & \textbf{625.8 / 99} \\
gobmk-O2 & 6 & 1337.7 / 100 & 1196 / 99.1 & \textbf{815.1 / 99} \\
gobmk-O3 & 6.4 & 1416.2 / 100 & 1184 / 99.1 & \textbf{1001.5 / 98} \\
\hline
perlbench-O0 & 2.8 & 400.3 / 99.1 & 202.7 / 99.1 & \textbf{182.1/ 99} \\
perlbench-O1 & 3.9 & 379.7 / 100 & 356.9 / 100 & \textbf{170.1 / 96} \\
perlbench-O2 & 4.1 & 377.6 / 100 & 343.1 / 100 & \textbf{192 / 95} \\
perlbench-O3 & 4.9 & 453.4 / 100 & 440.8 / 100 & \textbf{291.8 / 98} \\
\hline
gcc-O0 & 8.7  & 534.8 / 99.3 & 486.3 / 99.3 & \textbf{241.9 / 99} \\
gcc-O1 & 11.9 & 491 / 99.5 & 478.5 / 99.5 & \textbf{357.8 / 98} \\
gcc-O2 & 12.1 & 427.7 / 99.5 & 397.7 / 99.5 & \textbf{388.8 / 99} \\
gcc-O3 & 14.3 & 544.3 / 99.6 & 503.4 / 99.6 & \textbf{474.9 / 99} \\
\hline
thttpd-O0 & 0.2 & 8 / 100 & 5 / 100 & \textbf{3 / 100} \\
thttpd-O1 & 0.2 & 8 / 100 & \textbf{5 / 100} & \textbf{5 / 100} \\
thttpd-O2 & 0.2 & 14 / 100 & 4 / 100 & \textbf{3 / 100} \\
thttpd-O3 & 0.3 & 14 / 100 & \textbf{1 / 100} & \textbf{1 / 100} \\
\hline
memcached-O0 & 0.4 & 1 / 92 & \textbf{0.9 / 92} & \textbf{0.9 / 92} \\
memcached-O1 & 0.5 & 0.9 / 71 & \textbf{0.7 / 71} & \textbf{0.7 / 71} \\
memcached-O2 & 0.6 & 1.4 / 70 & 0.9 / 70 & \textbf{0.8 / 70} \\
memcached-O3 & 0.6 & 1.4 / 71 & 0.9 / 71 & \textbf{0.8 / 71} \\
\hline
lighttpd-O0 & 0.8 & 35.5 / 100 & 9.7 / 100 & \textbf{7.7 / 100} \\
lighttpd-O1 & 1.1 & 35.5 / 100 & 11.7 / 100 & \textbf{11 / 100} \\
lighttpd-O2 & 1.2 & 34.8 / 100 & 15.4 / 100 & \textbf{14.6 / 100} \\
lighttpd-O3 & 1.3 & 35.1 / 100 & \textbf{15.3 / 100} & \textbf{15.3 / 100} \\
\hline
exim-O0 & 3.4 & 31.1 / 100 & 11.3 / 100 & \textbf{7.8 / 99} \\
exim-O1 & 4.1 & 29.6 / 100 & 12.5 / 100 & \textbf{10.6 / 100} \\
exim-O2 & 4.3 & 30.6 / 100 & 12.6 / 100 & \textbf{10.5 / 100} \\
exim-O3 & 4.5 & 40.6 / 100 & 18.2 / 100 & \textbf{12.7 / 100} \\
\hline
nginx-O0 & 4.4 & 444 / 100 & 306.1 / 100 & \textbf{143.6 / 100} \\
nginx-O1 & 5.1 & 463.4 / 100 & 305.8 / 100 & \textbf{160.4 / 99} \\
nginx-O2 & 5.2 & 527.3 / 100 & 359.1 / 99.9 & \textbf{183.6 / 99} \\
nginx-O3 & 5.4 & 515.1 / 100 & 351.2 / 99.9 & \textbf{176.9 / 99} \\
\hline
Average & 2.7 &  195.9 / 98.2 & 158.3 / 98.2 & \textbf{109.2 / 97.8} \\
\hline
\end{tabular}
\captionsetup{justification=justified,singlelinecheck=false}
\caption*{\scriptsize Each cell reports \texttt{AICT / AICT\_Recall}. Binary size is shown in MB. Lower AICT with similar recall indicates improved precision.}

\label{tab:layerwise-results}
\end{table}

\subsection{RQ3: Comparison with Existing Solutions} \label{sec:rq3}
\textbf{Academic Tools:} To evaluate the effectiveness of our framework, \sysname{}, we compare it against three representative academic baselines: CALLEE~\cite{callee}, AttnCall~\cite{Attncall}, and BinDSA~\cite{bindsa}. We exclude BPA~\cite{bpa} since it forms the foundational static engine (Layer 1) of \sysname{} itself. SchedExec~\cite{SchedExec} is also omitted, as it excludes key benchmarks like perlbench, gcc, and other real-world binaries, limiting its completeness and generalizability.

\begin{table}[t]
\centering
\tiny
\setlength{\tabcolsep}{2pt}
\caption{Configurations of \sysname{} under different pruning regimes and evaluation objectives.}

\label{tab:iresolvex-modes-merged}
\begin{tabular}{@{}lcccc@{}}
\toprule
\textbf{Mode} & \textbf{Regime} & \textbf{L2b $\Delta$} & \textbf{p-IndirectCFG $\tau$} & \textbf{Focus} \\
\midrule
\textbf{\sysname{}@C-R} & Conservative & $\pm0.1$ & Trace-guided (Recall-tuned) & completeness-oriented \\
\textbf{\sysname{}@C-F} & Conservative & $\pm0.1$ & Trace-guided (F1-tuned) & balanced pruning \\
\textbf{\sysname{}@A-R} & Aggressive   & $\pm0.5$ & Trace-guided (Recall-tuned) & stronger pruning \\
\textbf{\sysname{}@A-F} & Aggressive   & $\pm0.5$ & Trace-guided (F1-tuned) & high-confidence pruning \\
\bottomrule
\end{tabular}
\captionsetup{justification=justified,singlelinecheck=false}
\caption*{\scriptsize $\Delta$: L2b score adjustment controlling the extent of learning-based refinement (larger $\Delta$ means stronger pruning). 
$\tau$: pIndirectCFG edge-inclusion threshold, trace-guided when dynamic traces are available or static otherwise—lower for recall-preserving and moderate for F1-preserving modes. Conservative regimes favor completeness; aggressive ones emphasize keeping high-confidence edges.}
\end{table}

\textbf{Evaluation Parameters and Configurations.}  
We primarily use AICT recall in RQ1 to evaluate pruning effectiveness while maintaining completeness. To accommodate different analysis objectives, \sysname{} supports four evaluation modes that vary along two complementary dimensions: pruning aggressiveness and preservation goal.
As summarized in Table~\ref{tab:iresolvex-modes-merged}, the conservative mode applies gentler L2b refinement to maintain completeness, while the aggressive mode readjusts L2a’s scores more strongly to suppress over-approximated targets.
Independently, each mode can prioritize either recall preservation---favoring comprehensive coverage of potential targets, or F1 balance---favoring reduced false positives. Note that we omit the case that pursues best precision, because we believe it is meaningless to pursue precision without maintaining a reasonable recall rate. The balance between the two objectives is achieved by selecting the pIndirectCFG edge-inclusion threshold accordingly. 
Together, these configurations illustrate how \sysname{} can be used for distinct analysis contexts: recall-preserving modes suit security-critical tasks where completeness is essential, while F1-preserving modes favor analyses seeking compact, high-confidence control-flow edges.
F1 is computed per callsite (Appendix~\ref{appendix:global_recall}).

\textbf{Fair Metric Comparison and Reproducibility Barriers.} The evaluation metrics, recall and F1 scores, rely on dynamic trace coverage. To ensure fair comparison, we follow the same SPEC-provided input-based tracing strategy as AttnCall and BinDSA, making reported metrics comparable. However, since dynamic targets capture only a subset of the ground truth, recall tends to be overestimated and precision underestimated, as noted in prior works~\cite{Attncall, bindsa, bpa}.

However, CALLEE’s setup diverges from other tools, complicating the recall comparisons. While it released dynamic ground truth, the binaries were missing, and artifacts lacked the embedding model, making reproduction infeasible. Even after a best-effort re-engineering of CALLEE based on their paper, our reevaluation with our dynamic data, reported only 50\% AICT recall and 70\% global recall—far below the claimed 90.9\% global recall. Moreover, CALLEE’s reliance on IDA’s DataXRef under-approximates address-taken functions—detecting only 414 in gcc, while \sysname{} recovers 1{,}207 on average\footnote{Although CALLEE reports slightly higher address-taken counts, its results remain under-approximated; our tests on x86/x64 builds confirm IDA DataXRef’s unreliability. Since CALLEE’s binaries were not released, we could not verify its claims.}—leading to 417K missing pairs. Re-running its fuzzing setup~\cite{ibresolver2025} confirmed that 33.4\% of valid callees were missed. Similar but less severe patterns appear for perlbench and gobmk (see Table~\ref{tab:addr_taken_comparison} in Appendix~\ref{appendix:rq3-addr}). Overall, these issues highlight CALLEE’s scalability and completeness limitations.

Moreover, considering CALLEE used the same \texttt{gcc-9} compiled binaries and compared against BPA, its fuzzing-based dynamic data should, in principle, yield a comparable or higher number of callsite–callee pairs. It recorded only 187 callsite--callee pairs per benchmark on average vs our 353. Dynamic ground truth for benchmarks like perlbench, milc, and sphinx3 as well as real-world binaries were absent from their released dataset. 
This 47\% gap arises from (1) unhandled command-line arguments and (2) shallow input exploration in the naive fuzzing set-up. Assuming recall scales linearly with trace size, its true recall may fall as low as:
\[
0.53 \times R_{\text{obs}} \le R_{\text{true}} \le R_{\text{obs}} \quad \Rightarrow \quad 42.3\% \le R_{\text{true}} \le 90.9\%.
\]
However, we acknowledge that achieving high dynamic coverage remains a challenge across all works, highlighting the need for program-aware, grammar-based fuzzing.  Nevertheless, for fairness, we adhered to the same experimental setup as AttnCall and BinDSA. In addition, while we include CALLEE in our comparison, its recall and F1 scores are not directly comparable due to different dynamic data setting. Since AICT is independent of dynamic traces, we evaluate CALLEE on that metric only.  We also ignore its missing callsite--callee pairs caused by IDA DataXRef---including them would further favor our results. AttnCall and BinDSA also face reproducibility issues, AttnCall's code lacks a trained model, and BinDSA is unreleased. Hence, We rely solely on the reported numbers from CALLEE, AttnCall, and BinDSA. As none of these works specify optimization levels of their evaluation binaries, we align AICT baselines with BPA: O2 binaries for CALLEE and BinDSA, and O0 for AttnCall.

\begin{table}[t]
\centering
\tiny
\caption{Comparing AICT of \sysname{} across different configurations with state-of-the-art baselines.}
\label{tab:aict-compare-sota}
\setlength{\tabcolsep}{3pt}   
\renewcommand{\arraystretch}{1.1}
\resizebox{\linewidth}{!}{%
\begin{tabular}{l|rr|rr|r|r|r}
\hline
Benchmark & C-R & C-F & A-R & A-F & CALLEE & AttnCall & BinDSA \\
\hline
bzip2-O0   & 1  & 0.25 & 1  & 1  & –   & 1   & – \\
bzip2-O2   & 1.55 & 0.25 & 2  & 1.20 & 1.4 & –  & 1 \\
\hline
sjeng-O0   & 6  & 6  & 6  & 6  & – & 6 & – \\
sjeng-O2   & 6  & 6  & 7  & 7  & 7 & – & 7 \\
\hline
milc-O0    & 2  & 2  & 2  & 2  & – & 3 & – \\
milc-O2    & 2  & 2  & 2  & 2  & 2 & – & 1 \\
\hline
sphinx3-O0 & 0.75 & 0.75 & 1.25 & 1.25 & – & 5.7 & – \\
sphinx3-O2 & 0.57 & 0.57 & 0.57 & 0.57 & 5.6 & – & – \\
\hline
hmmer-O0   & 2.40 & 1.44 & 2.78 & 1.44 & – & 6.6 & – \\
hmmer-O2   & 2.30 & 2.30 & 1.70 & 1.70 & 7.2 & – & 3.3 \\
\hline
h264ref-O0 & 3.41 & 1.10 & 2.40 & 0.03 & – & 21.5 & – \\
h264ref-O2 & 5.90 & 0.02 & 1  & 0.06 & 20.9 & – & 2.04 \\
\hline
gobmk-O0        & 532.90 & 170.18 & 485.48 & 130.52 & – & 422 & – \\
gobmk-O2        & 815.07 & 21.91  & 564.43 & 151.89 & 672.4 & – & 86.1 \\
\hline
perlbench-O0    & 182.10 & 1.06   & 166.53 & 2.78  & – & 313.2 & – \\
perlbench-O2    & 192  & 16.36  & 148.50 & 39.15 & 354 & – & 20.8 \\
\hline
gcc-O0      & 241.90 & 5.45 & 209.40 & 6.92 & – & 274 & – \\
gcc-O2      & 388.81 & 0.73 & 113.98 & 5.90 & 338 & – & 111.2 \\
\hline
thttpd-O0   & 3  & 3  & 3  & 3  & – & 7 & – \\
thttpd-O2   & 3  & 3  & 3  & 3  & – & – & – \\
\hline
memcached-O0  & 0.89 & 0.89 & 0.93 & 0.93 & – & 2.1 & – \\
memcached-O2  & 0.79 & 0.60 & 0.71 & 0.71 & 11.3 & – & 1.2 \\
\hline
lighttpd-O0   & 7.68 & 0.02 & 7.66 & 0.22 & – & 17.4 & – \\
lighttpd-O2   & 14.61 & 0.01 & 7.20 & 0.20 & 31.7 & – & 17.6 \\
\hline
exim-O0     & 7.81 & 0.19 & 6.17 & 0.45 & – & 15.1 & – \\
exim-O2     & 10.51 & 0.08 & 8.78 & 0.28 & 22.4 & – & 5 \\
\hline
nginx-O0    & 143.59 & 0.18 & 131.86 & 2.42 & – & 296.5 & – \\
nginx-O2    & 183.60 & 0.62 & 170  & 4.09 & 383 & – & 8.7 \\
\hline
average-O0   & 81.10 & 13.80 & 73.30 & 11.40 & – & 99.36 & – \\
\hline
average-O2   & 116.20 & 3.90 & 73.60 & 15.60 & 142.84 & – & 22.08 \\
\hline

\end{tabular}%

}
\captionsetup{justification=justified,singlelinecheck=false}
\caption*{\footnotesize 
C-R = {\sysname{}}@C-R (Conservative Recall-Preserving), 
C-F = {\sysname{}}@C-F (Conservative F1-Preserving), 
A-R = {\sysname{}}@A-R (Aggressive Recall-Preserving), 
A-F = {\sysname{}}@A-F (Aggressive F1-Preserving). 
Each cell reports \texttt{AICT}.}
\end{table}

\begin{table}[t]
\centering
\tiny
\caption{Comparing Recall / F1 of \sysname{} across different configurations with state-of-the-art baselines on SPEC2006 Benchmark.}
\label{tab:aict-compare-sota-recall}
\setlength{\tabcolsep}{3pt}   
\renewcommand{\arraystretch}{1.1}
\resizebox{\linewidth}{!}{%
\begin{tabular}{l|rr|rr|r|r|r}
\hline
Benchmark & C-R & C-F & A-R & A-F & CALLEE & AttnCall & BinDSA \\
  & (Recall) & (F1) & (Recall) & (F1) &  & (Recall) & (F1) \\
\hline
bzip2-O0   & 100 & 79 & 100 & 75 & – & 100 & – \\
bzip2-O2   & 100 & 79 & 100 & 72 & – & – & 100 \\
\hline
sjeng-O0   & 100 & 100 & 100 & 100 & – & 100 & – \\
sjeng-O2   & 100 & 93 & 100 & 93 & – & – & 92.5 \\
\hline
milc-O0      & 100 & 100 & 100 & 100 & – & 100 & – \\
milc-O2      & 100 & 100 & 100 & 100 & – & – & 100 \\
\hline
sphinx3-O0   & 100 & 91 & 100 & 88.9 & – & 100 & – \\
sphinx3-O2   & 100 & 94.3 & 100 & 94.3 & – & – & NA \\
\hline
hmmer-O0    & 100 & 89.8 & 100 & 89.8 & – & 100 & – \\
hmmer-O2    & 100 & 95.1 & 100 & 95.4 & – & – & 95.3 \\
\hline
h264ref-O0  & 99.8 & 89.8 & 99.3 & 87.8 & – & 92.3 & – \\
h264ref-O2  & 99.7 & 98.9 & 99 & 97.1 & – & – & 78.8 \\
\hline
gobmk-O0        & 98 & 55.7 & 96 & 64.7 & – & 96.9 & – \\
gobmk-O2        & 99 & 42 & 86 & 53.4 & – & – & 70.5 \\
\hline
perlbench-O0    & 99 & 55.6 & 93 & 65.3 & – & 78.4 & – \\
perlbench-O2    & 95 & 56.3 & 85 & 57 & – & – & 65.6 \\
\hline
gcc-O0       & 99 & 70.8 & 98.5 & 70.2 & – & 93.2 & – \\
gcc-O2       & 99 & 70.7 & 90 & 68.6 & – & – & 54.5 \\
\hline
average-O0   & 99.5 & 81.3 & 98.5 & 82.4 & – & 95.6 & – \\
\hline
average-O2   & 99.2 & 81 & 95.6 & 81.2 & – & – & 82.1 \\
\hline

\end{tabular}%

}
\captionsetup{justification=justified,singlelinecheck=false}
\caption*{\scriptsize 
C-R = {\sysname{}}@C-R (Conservative Recall-Preserving), 
C-F = {\sysname{}}@C-F (Conservative F1-Preserving), 
A-R = {\sysname{}}@A-R (Aggressive Recall-Preserving), 
A-F = {\sysname{}}@A-F (Aggressive F1-Preserving). 
Each cell reports \texttt{AICT\_Recall} for recall-preserving and \texttt{AICT\_F1} for F1-preserving configurations. 
CALLEE reports no benchmark-specific numbers; for AttnCall, each cell reports \texttt{AICT\_Recall}, and for BinDSA, each cell reports \texttt{AICT\_F1}.}
\end{table}

\textbf{Comparison with State-of-the-Art.} For clarity, we compare \sysname{} to state-of-the-art baselines using only \texttt{O0} and \texttt{O2} binaries, though results generalize to other levels. AICT comparisons (dynamic-trace–independent) appear in Table~\ref{tab:aict-compare-sota}, and recall/F1 comparisons in Table~\ref{tab:aict-compare-sota-recall}. Since prior work collects dynamic traces only for SPEC, Table~\ref{tab:aict-compare-sota-recall} covers SPEC.

As shown in Table~\ref{tab:aict-compare-sota}, \sysname{} delivers consistent AICT reductions across all evaluation modes. In conservative mode, \sysname{}@C-R achieves \textbf{18\% more reduction} than both CALLEE and AttnCall. In the F1-preserving setting (\sysname{}@C-F), it improves over CALLEE, BinDSA, and AttnCall by \textbf{97\%, 82\%, and 86.2\%}, respectively. In aggressive mode, \sysname{}@A-R outperforms CALLEE and AttnCall by \textbf{48.5\% and 26.2\%}, while \sysname{}@A-F improves over CALLEE, BinDSA, and AttnCall by \textbf{89.2\%, 30\%, and 88.6\%}, respectively. Importantly, recall remains high in recall-preserving modes: 98.1\% for \sysname{}@C-R and 96.4\% for \sysname{}@A-R. Even in F1-preserving modes, \sysname{} sustains strong recall---82.6\% for \sysname{}@C-F and 88.8\% for \sysname{}@A-F---while maintaining 80\% F1 and precision on average.\footnote{Recall/F1/Precision are averaged from the full evaluation logs; Table~\ref{tab:aict-compare-sota} does not list them, and Table~\ref{tab:aict-compare-sota-recall} shows recall rates for part of the SPEC.}

While BinDSA reports lower AICT under recall-preserving settings, it suffers from substantial incompleteness due to imprecise data-structure recovery. Even for small binaries like sphinx3, it resolves no targets despite dynamic evidence, and it misses 18–37\% of valid callsites in larger programs (perlbench, gcc, nginx, lighttpd). In contrast, \sysname{} offers configurable precision–coverage trade-offs, supporting both high-coverage and high-precision scenarios. As shown in Table~\ref{tab:aict-compare-sota-recall}, on SPEC2006 \sysname{} attains higher recall in recall-preserving configurations and comparable F1, while still achieving substantial AICT reduction.

\textbf{RE Tools:} We observe a significant gap between mainstream RE tools and academic approaches in resolving indirect calls. On average, IDA and GHIDRA achieve low AICT (1.8 and 0.8) and recall (16\% and 11.8\%), with GHIDRA introducing false positives. Angr's CFGEmulated attains better recall (31.4\%) but remains impractically slow, timing out on larger binaries (Table~\ref{tab:tool_aict_recall}). These results show a need for a flexible solution that connects academic and RE tools. p-IndirectCFG addresses this gap.

\begin{table}[t!]
\centering
\scriptsize
\caption{Comparison of RE Tools}
\label{tab:tool_aict_recall}
\begin{tabular}{|l|c|c|c|c|}
\hline
\textbf{Binary} & \textbf{IDA} & \textbf{GHIDRA} & \multicolumn{2}{c|}{\textbf{Angr}} \\ 
\cline{4-5}
                  &              &                 &  CFGEmulated &  CFGFast \\
\hline
bzip2     & 0 / 0     & 0.9 / 25    & 0.1 / 5    & 0 / 0     \\
sjeng     & 7 / 100   & 0 / 0       & 1 / 0    & 0 / 0     \\
milc      & 0 / 0     & 0 / 0       & 1 / 100    & 0 / 0     \\
sphinx3   & 0 / 0     & 0 / 0       & 0 / 0      & 0 / 0     \\
hmmer     & 0 / 0         & 0 / 0        & 0.9 / 87.4  & 0 / 0     \\
h264ref   & 0 / 0         & 2.1 / 8.3    & 0.9 / 1.2   & 0 / 0     \\
gobmk     & 0 / 0         & 1 / 26.6     & 0.5 / 27.8  & 0 / 0     \\
perlbench & 7.4 / 32.1    & 0.9 / 30.1   & 14.1 / 29.3 & 0.3 / 26  \\
gcc       & 1.7 / 11.9    & 2.5 / 16.4   & \textit{timeout }    & 0.2 / 11.8 \\
\hline
average       & 1.8 / 16      & 0.8 / 11.8   & \textbf{1.9 / 31.4}  & 0.1 / 4.2 \\

\hline
\end{tabular}
\captionsetup{justification=justified,singlelinecheck=false}
\caption*{\scriptsize Each cell reports \texttt{AICT / AICT\_Recall}}

\end{table}

\begin{figure}[t!]
    \centering
    \includegraphics[width=0.59\linewidth]{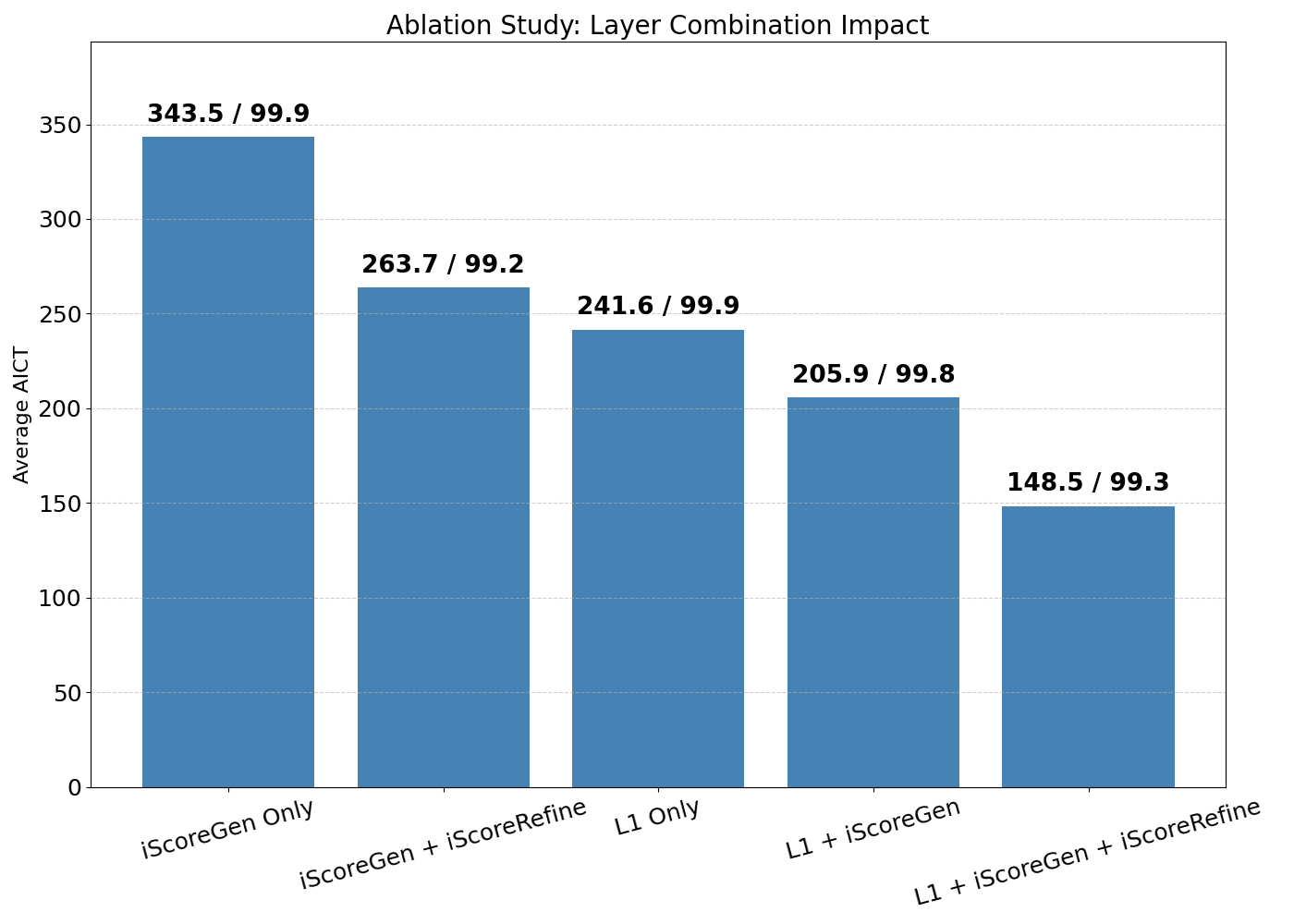}
    \caption{Ablation study with average \texttt{AICT} and \texttt{AICT\_recall} annotated on each bar.}
    \label{fig:ablation-study}
\end{figure}

\subsection{RQ4: Ablation Study on Layer Combinations}
To assess the contributions of individual components within \sysname{}, we perform an ablation study over all meaningful permutations of its three layers: L1 (BPA), L2a (iScoreGen), and L2b (iScoreRefine). The five valid combinations that isolate the effect of SA and ML are: (Figure~\ref{fig:ablation-study}):  (1) L1 only (BPA), (2) L1+iScoreGen (L1+L2a), (3) full pipeline L1+iScoreGen+iScoreRefine (L1+L2a+L2b), (4) iScoreGen only (L2a), and (5) iScoreGen+iScoreRefine (L2a+L2b). Note that L1+L2b is invalid, as L2b relies on L2a's outputs. For this study, we report results under the conservative, recall-preserving configuration (\sysname{}@C-R); other configurations exhibit similar results. The results show that L1 provides strong recall but suffers from over-approximation. Adding iScoreGen (L2a) significantly improves precision, and iScoreRefine (L2b) further reduces false positives through static inter-procedural validation. In contrast, ML-only configurations (L2a and L2a+L2b) inflate target sets due to lack of a conservative base. Overall, the best results emerge from combining all layers: L1 anchors soundness, L2a improves precision, and L2b enforces contextual pruning.

\subsection{RQ5: p-IndirectCFG Thresholds and its Effects}
\begin{figure}[t!]
    \centering
    \includegraphics[width=\linewidth]{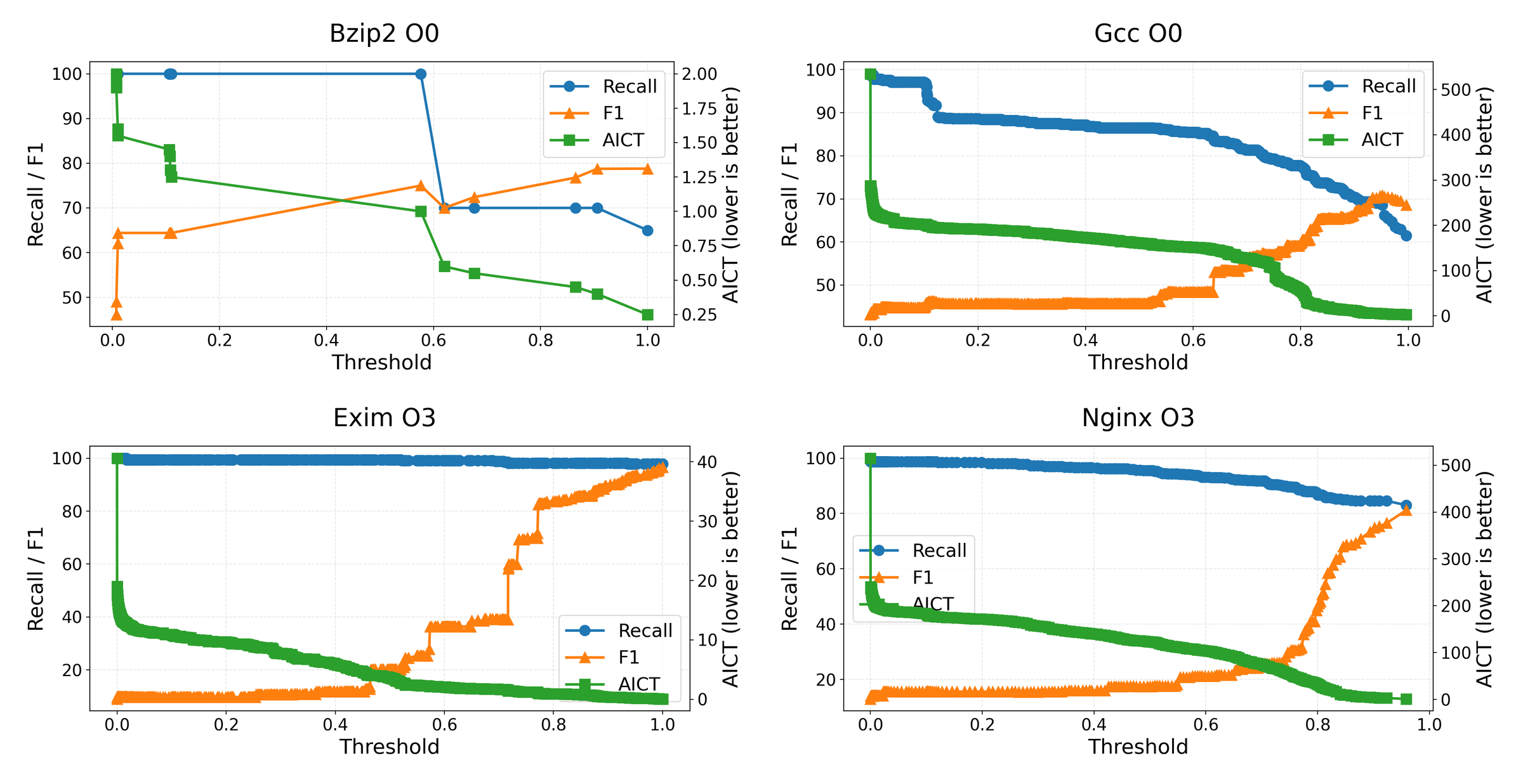}
    \caption{Threshold sensitivity analysis showing variation of Recall, F1, and AICT with threshold. Left axis: \texttt{AICT\_Recall or AICT\_F1}; right axis: AICT.}
    \label{fig:threshold_sensitivity}
\end{figure}

We study how threshold selection in p-IndirectCFG affects evaluation metrics through a sensitivity analysis on randomly selected benchmarks varying in optimization and binary size—covering the smallest (bzip2), largest (gcc), and real-world O3 binaries (Figure \ref{fig:threshold_sensitivity}). For this study, we report results under the conservative, recall-preserving configuration (\sysname{}@C-R); other configurations exhibit similar results. A consistent trend emerges: increasing the threshold consistently decreases AICT and recall while improving F1. Although the rate of change varies across benchmarks. This trend is consistently observed across other benchmarks in our evaluation set. Notably, all exhibit an initial sharp AICT drop; smaller binaries like bzip2 show immediate F1 gains, while larger ones like gcc improve later as more false positives are pruned. This adjustable threshold allows p-IndirectCFG to adapt to different analysis needs: low thresholds preserve recall-critical edges, while high thresholds favor precision by pruning spurious paths. As shown in our case studies (Appendix~\ref{appendix:case_study}), this adaptability helps retain aliasing-driven edges critical to malicious-payload reachability, while high thresholds eliminate BPA-induced self-loops and structurally invalid paths.
\section{Discussion and Limitations}
\textbf{L2a Fine-tuning and Design Choices:} L2a learns a function-signature-guided feature space that generalizes from direct to indirect calls, reducing reliance on large-scale dynamic tracing—validated by our evaluation. To assess fine-tuning benefits, we fine-tuned the L2a model with an additional 10k dynamically extracted indirect callsite–callee pairs. This yielded a 1.5\% AICT drop with no recall loss, indicating that fine-tuning helps capture rare patterns. However, when L2b is applied on top of the fine-tuned L2a, the gain reduces to 0.8\%, as backward reasoning already subsumes much of this benefit. While additional dynamic data may offer incremental gains, we expect diminishing returns once L2b is included. Moreover, replacing L2a’s DNN with a more expressive model like a transformer may improve pruning, which we leave to future work. L2a's Algorithm~\ref{alg:callee-feature} also includes compiler-specific x86 heuristics from our observations to improve Angr’s unreliable calling convention analysis. Since Angr already relies on heuristics, we enhanced them for stability—though stronger static frontends could eliminate this need.

\textbf{Practicality \& Scalability:} We evaluated the runtime of each layer—L1 (static analysis), L2a (feature extraction + inference), L2b (backward reasoning), and dynamic tracing. The most expensive components are L1 and dynamic tracing, but their parallelizability limits end-to-end overhead. Compared to Angr’s high-recall CFGEmulated, \sysname{} is up to 74\% faster on small binaries (avg. $<$100s at 100\% recall). For large binaries like gcc, where Angr times out ($>$24h), \sysname{} completes in ~10h at 98.8\% recall—the highest runtime in our suite.  Although slower than pure ML-based approaches, \sysname{} offers capabilities they lack, including static, inter-procedural reasoning and threshold-based flexibility for recall-versus precision-sensitive pruning.


\textbf{ISA and Language Support:} 
\sysname{} currently targets C-based x86-32 binaries due to limitations of the BPA engine, which supports only this configuration. While L1 depends on BPA and inherits this constraint, it is theoretically architecture- and language-agnostic. In contrast, Layers 2a and 2b are decoupled from both x86-32 and C, and can operate on any static-analysis-driven CFG. As shown in our ablation (Figure~\ref{fig:ablation-study}), \sysname{} can run without L1 by treating address-taken functions as possible targets and applying L2 pruning, achieving 23.2\% AICT reduction---close to the 38.5\% reduction when using L1+L2.
To support x86-64, L1 could be adapted using a sound alternative such as TypeArmor~\cite{typearmour}, which offers similar over-approximation. Notably, \sysname{}'s L2 does not require a sound L1, i.e., L2 can operate independently on CFGs from any tool, including CALLEE and AttnCall, ranking or pruning indirect edges post hoc. This makes \sysname{} broadly applicable across toolchains.  In contrast, existing ML tools are tightly coupled to x86-64 because they rely on calling-convention-register-specific slicing logic, making x86-32 support non-trivial. Extending support to x86-64 and ARM via pluggable front-ends is a key future direction.

\textbf{Score Adjustment in iScoreRefine (L2b):}
L2b currently uses a static score to adjust L2a’s output, offering flexibility for different analysis goals. While effective, this method does not account for program-specific factors such as indirect callsite density, callee ambiguity, or control-flow complexity. To improve adaptability, we plan to replace thresholding with a heterogeneous Graph Neural Network (GNN)~\cite{gnn1, gnn2} that learns to adjust scores using inter-procedural structure and semantic context. The GNN operates on a multi-relational graph of callsites, candidate callees, memory locations, and instructions, with edges capturing control and data flow. Nodes are enriched with subgraphs capturing argument flows, dereferences, and cross-function control patterns. We use iScoreGen’s compatibility scores as weak supervision to enable unsupervised refinement. Preliminary results on a subset of benchmarks show that this learning-based refinement yields additional AICT reductions without loss in recall, validating its potential. We leave full integration and evaluation to future work.
\textbf{Sparsity in Dynamic Tracing for Per-benchmark Threshold Selection:} Indirect call behavior varies widely across programs, making a universal dynamic profile infeasible. p-IndirectCFG instead uses optional partial traces (30\% of total), avoiding reliance on exhaustive profiling. As future work, we aim to explore lightweight RL~\cite{rl, rl1} to automate pruning—using forced-execution traces only during training and enabling static, policy-driven thresholds at test time, where RL naturally captures precision–recall trade-offs as a reward-driven optimization problem.

\section{Conclusion}
We introduced \sysname{}, a hybrid framework that resolves indirect call targets by combining value-set analysis, ML-based signature scoring, and selective inter-procedural reasoning. This design balances soundness and precision, separating generalizable signatures from context-specific behaviors without requiring dynamic supervision. Our p-IndirectCFG further bridges the gap between academic research and RE tools. Empirically, \sysname{} reduces AICT while maintaining high recall, outperforming state-of-the-art approaches. As a flexible and modular framework, \sysname{} is well-equipped for a variety of downstream tasks that require controllable precision-recall trade-offs, fulfilling a critical gap in the current literature.


\section{Ethics considerations} 
This work improves indirect call resolution on self-compiled SPEC and open-source binaries only, with no personal data or human subjects, benefiting open-source maintainers, security analysts, and end users via more accurate CFGs and easier hardening. Although reverse engineering is dual-use, we release artifacts with guidance for authorized use only and log only control-flow metadata, so we expect defensive benefits to outweigh potential misuse.

\bibliographystyle{plain}
\bibliography{main}

@InProceedings{bpa,
  Title                    = {Refining Indirect Call Targets at the Binary Level},
  Author                   = {Sun Hyoung Kim and Cong Sun and Dongrui Zeng and Gang Tan},
  Booktitle                = {Network and Distributed System Security Symposium (NDSS)},
  Year                     = {2021}
}

@inproceedings{typearmour,
author = {van der Veen, Victor and Göktaş, Enes and Contag, Moritz and Pawloski, Andre and Chen, Xi and Rawat, Sanjay and Bos, Herbert and Holz, Thorsten and Athanasopoulos, Elias and Giuffrida, Cristiano},
year = {2016},
month = {05},
pages = {},
title = {A Tough call: Mitigating Advanced Code-Reuse Attacks At The Binary Level},
doi = {10.1109/SP.2016.60}
}

@InProceedings{SchedExec,
author="Shi, Yangyang
and Tian, Linan
and Chen, Liwei
and Yang, Yanqi
and Shi, Gang",
editor="Garcia-Alfaro, Joaquin
and Kozik, Rafa{\l}
and Chora{\'{s}}, Micha{\l}
and Katsikas, Sokratis",
title="Scheduled Execution-Based Binary Indirect Call Targets Refinement",
booktitle="Computer Security -- ESORICS 2024",
year="2024",
publisher="Springer Nature Switzerland",
address="Cham",
pages="3--23",
isbn="978-3-031-70896-1"
}

@inproceedings{disa,
  author       = {Peicheng Wang and
                  Monika Santra and
                  Mingyu Liu and
                  Cong Sun and
                  Dongrui Zeng and
                  Gang Tan},
  title        = {Disa: Accurate Learning-based Static Disassembly with Attentions},
  booktitle = {31st ACM Conference on Computer and Communications Security ({CCS})},
  year      = {2025}
}

@misc{callee,
      title={Callee: Recovering Call Graphs for Binaries with Transfer and Contrastive Learning}, 
      author={Wenyu Zhu and Zhiyao Feng and Zihan Zhang and Jianjun Chen and Zhijian Ou and Min Yang and Chao Zhang},
      year={2022},
      eprint={2111.01415},
      archivePrefix={arXiv},
      primaryClass={cs.SE},
      url={https://arxiv.org/abs/2111.01415}, 
}

@InProceedings{Attncall,
author="Sun, Rui
and Guo, Yinggang
and Wang, Zicheng
and Zeng, Qingkai",
editor="Tsudik, Gene
and Conti, Mauro
and Liang, Kaitai
and Smaragdakis, Georgios",
title="AttnCall: Refining Indirect Call Targets in Binaries with Attention",
booktitle="Computer Security -- ESORICS 2023",
year="2024",
publisher="Springer Nature Switzerland",
address="Cham",
pages="391--409",
isbn="978-3-031-51482-1"
}

@article{bindsa,
author = {Gao, Lian and Yin, Heng},
title = {BinDSA: Efficient, Precise Binary-Level Pointer Analysis with Context-Sensitive Heap Reconstruction},
year = {2025},
issue_date = {July 2025},
publisher = {Association for Computing Machinery},
address = {New York, NY, USA},
volume = {2},
number = {ISSTA},
url = {https://doi.org/10.1145/3728928},
doi = {10.1145/3728928},
journal = {Proc. ACM Softw. Eng.},
month = jun,
articleno = {ISSTA053},
numpages = {22}
}

@misc{idapro,
  author = {{Hex-Rays}},
  title = {{IDA Pro Disassembler and Debugger}},
  howpublished = {\url{https://hex-rays.com/ida-pro/}},
  note = {Version 7.5, Accessed: 2025-07-29}
}

@misc{ghidra,
  author = {{National Security Agency}},
  title = {{Ghidra Software Reverse Engineering Framework}},
  howpublished = {\url{https://ghidra-sre.org/}},
  note = {Accessed: 2025-07-29}
}

@inproceedings{angr,
  title={{SoK}: (State of) The Art of Symbolic Execution with the {angr} Framework},
  author={Shoshitaishvili, Yan and Wang, Ruoyu and Salls, Christopher and Stephens, Nick and Grosen, Mario and Dutcher, Sophia and Feng, John and Hauser, Christophe and Kruegel, Christopher and Vigna, Giovanni},
  booktitle={IEEE Symposium on Security and Privacy (S\&P)},
  pages={89--104},
  year={2016},
  organization={IEEE}
}

@inproceedings{analysis1,
	title = {A {Tough} {Call}: {Mitigating} {Advanced} {Code}-{Reuse} {Attacks} at the {Binary} {Level}},
	url = {Paper=https://download.vusec.net/papers/typearmor_sp16.pdf Code=https://github.com/vusec/typearmor},
	booktitle = {S\&{P}},
	author = {van der Veen, Victor and Goktas, Enes and Contag, Moritz and Pawlowski, Andre and Chen, Xi and Rawat, Sanjay and Bos, Herbert and Holz, Thorsten and Athanasopoulos, Elias and Giuffrida, Cristiano},
	month = may,
	year = {2016},
}

@inproceedings{analysis2,
author = {van der Veen, Victor and Andriesse, Dennis and G\"{o}kta\c{s}, Enes and Gras, Ben and Sambuc, Lionel and Slowinska, Asia and Bos, Herbert and Giuffrida, Cristiano},
title = {Practical Context-Sensitive CFI},
year = {2015},
isbn = {9781450338325},
publisher = {Association for Computing Machinery},
address = {New York, NY, USA},
url = {https://doi.org/10.1145/2810103.2813673},
doi = {10.1145/2810103.2813673},
booktitle = {Proceedings of the 22nd ACM SIGSAC Conference on Computer and Communications Security},
pages = {927–940},
numpages = {14},
location = {Denver, Colorado, USA},
series = {CCS '15}
}

@inproceedings{sa1,
  title={BinPointer: towards precise, sound, and scalable binary-level pointer analysis},
  author={Kim, Sun Hyoung and Zeng, Dongrui and Sun, Cong and Tan, Gang},
  booktitle={Proceedings of the 31st ACM SIGPLAN International Conference on Compiler Construction},
  pages={169--180},
  year={2022}
}

@inproceedings{sa2,
  title={DEEPTYPE: Refining Indirect Call Targets with Strong Multi-layer Type Analysis},
  author={Tianrou Xia and Hong Hu and Dinghao Wu},
  booktitle={USENIX Security Symposium},
  year={2024},
  url={https://api.semanticscholar.org/CorpusID:271040295}
}

@inproceedings{sa3,
  title={Typro: Forward cfi for c-style indirect function calls using type propagation},
  author={Bauer, Markus and Grishchenko, Ilya and Rossow, Christian},
  booktitle={Proceedings of the 38th Annual Computer Security Applications Conference},
  pages={346--360},
  year={2022}
}

@inproceedings{dyn1,
author = {Li, Yuan and Wang, Mingzhe and Zhang, Chao and Chen, Xingman and Yang, Songtao and Liu, Ying},
title = {Finding Cracks in Shields: On the Security of Control Flow Integrity Mechanisms},
year = {2020},
isbn = {9781450370899},
publisher = {Association for Computing Machinery},
address = {New York, NY, USA},
url = {https://doi.org/10.1145/3372297.3417867},
doi = {10.1145/3372297.3417867},
booktitle = {Proceedings of the 2020 ACM SIGSAC Conference on Computer and Communications Security},
pages = {1821–1835},
numpages = {15},
location = {Virtual Event, USA},
series = {CCS '20}
}

@inproceedings{dyn2,
  title={Typesqueezer: When static recovery of function signatures for binary executables meets dynamic analysis},
  author={Lin, Ziyi and Li, Jinku and Li, Bowen and Ma, Haoyu and Gao, Debin and Ma, Jianfeng},
  booktitle={Proceedings of the 2023 ACM SIGSAC Conference on Computer and Communications Security},
  pages={2725--2739},
  year={2023}
}

@inproceedings{analysis3,
author = {Qiang, Weizhong and Huang, Yingda and Zou, Deqing and Jin, Hai and Wang, Shizhen and Sun, Guozhong},
year = {2017},
month = {05},
pages = {435-442},
title = {Fully Context-Sensitive CFI for COTS Binaries},
isbn = {978-3-319-59869-7},
doi = {10.1007/978-3-319-59870-3_28}
}

@InProceedings{vsa,
author="Balakrishnan, Gogul
and Reps, Thomas",
editor="Duesterwald, Evelyn",
title="Analyzing Memory Accesses in x86 Executables",
booktitle="Compiler Construction",
year="2004",
publisher="Springer Berlin Heidelberg",
address="Berlin, Heidelberg",
pages="5--23",
isbn="978-3-540-24723-4"
}

@inproceedings{souffle,
  title={Souffl{\'e}: On Synthesis of Program Analyzers},
  author={Herbert Jordan and Bernhard Scholz and Pavle Suboti{\'c}},
  booktitle={International Conference on Computer Aided Verification},
  year={2016},
  url={https://api.semanticscholar.org/CorpusID:7428346}
}

@inproceedings{aliaspointsto,
  title={Program Analysis and Specialization for the C Programming Language},
  author={Lars Ole Andersen and Peter Lee},
  year={2005},
  url={https://api.semanticscholar.org/CorpusID:20876553}
}

@misc{ibresolver2025,
  author = {{Immunant, Inc.}},
  title = {{ibresolver}: A QEMU TCG plugin for resolving indirect branches},
  year = {2025},
  howpublished = {\url{https://github.com/immunant/ibresolver}},
  note = {Accessed on July 29, 2025}
}

@inproceedings{gnn1,
  author       = {Thomas N. Kipf and
                  Max Welling},
  title        = {Semi-Supervised Classification with Graph Convolutional Networks},
  booktitle    = {5th International Conference on Learning Representations, {ICLR} 2017,
                  Toulon, France, April 24-26, 2017, Conference Track Proceedings},
  publisher    = {OpenReview.net},
  year         = {2017},
  url          = {https://openreview.net/forum?id=SJU4ayYgl},
  bibsource    = {dblp computer science bibliography, https://dblp.org}
}

@inproceedings{
gnn2,
title={Graph Attention Networks},
author={Petar Veličković and Guillem Cucurull and Arantxa Casanova and Adriana Romero and Pietro Liò and Yoshua Bengio},
booktitle={International Conference on Learning Representations},
year={2018},
url={https://openreview.net/forum?id=rJXMpikCZ},
}

@InProceedings{mibench,
  author={Guthaus, M.R. and Ringenberg, J.S. and Ernst, D. and Austin, T.M. and Mudge, T. and Brown, R.B.},
  booktitle={Proceedings of the Fourth Annual IEEE International Workshop on Workload Characterization. WWC-4 (Cat. No.01EX538)},
  title={MiBench: A free, commercially representative embedded benchmark suite},
  year={2001},
  volume={},
  number={},
  pages={3-14},
}

@misc{gnu_binutils,
  title = {The GNU Binutils},
  author = {{Free Software Foundation}},
  year = {2023},
  howpublished = {\url{https://www.gnu.org/software/binutils/}},
  note = {Accessed: 2025-10-03}
}

@misc{gnu_coreutils,
  title = {GNU Coreutils: The GNU Core Utilities},
  author = {{Free Software Foundation}},
  year = {2023},
  howpublished = {\url{https://www.gnu.org/software/coreutils/}},
  note = {Accessed: 2025-10-03}
}

@InProceedings{cfg_recovery,
author="Wang, Qianjin
and Li, Xiangdong
and Yue, Chong
and He, Yuchen",
editor="Zhang, Min
and Xu, Bin
and Hu, Fuyuan
and Lin, Junyu
and Song, Xianhua
and Lu, Zeguang",
title="A Survey of Control Flow Graph Recovery for Binary Code",
booktitle="Computer Applications",
year="2024",
publisher="Springer Nature Singapore",
address="Singapore",
pages="225--244",
isbn="978-981-99-8761-0"
}

@InProceedings{dirsym,
author="Ma, Kin-Keung
and Yit Phang, Khoo
and Foster, Jeffrey S.
and Hicks, Michael",
editor="Yahav, Eran",
title="Directed Symbolic Execution",
booktitle="Static Analysis",
year="2011",
publisher="Springer Berlin Heidelberg",
address="Berlin, Heidelberg",
pages="95--111",
isbn="978-3-642-23702-7"
}

@inproceedings{decomp,
author = {Yakdan, Khaled and Eschweiler, Sebastian and Gerhards-Padilla, Elmar and Smith, Matthew},
year = {2015},
month = {02},
pages = {},
title = {No More Gotos: Decompilation Using Pattern-Independent Control-Flow Structuring and Semantics-Preserving Transformations},
doi = {10.14722/ndss.2015.23185}
}

@inproceedings {fuzz2,
author = {Gwangmu Lee and Woochul Shim and Byoungyoung Lee},
title = {Constraint-guided Directed Greybox Fuzzing},
booktitle = {30th USENIX Security Symposium (USENIX Security 21)},
year = {2021},
isbn = {978-1-939133-24-3},
pages = {3559--3576},
url = {https://www.usenix.org/conference/usenixsecurity21/presentation/lee-gwangmu},
publisher = {USENIX Association},
month = aug
}

@InProceedings{fuzzsym,
author="Vin{\c{c}}ont, Ya{\"e}lle
and Bardin, S{\'e}bastien
and Marcozzi, Micha{\"e}l",
editor="A{\"i}meur, Esma
and Laurent, Maryline
and Yaich, Reda
and Dupont, Beno{\^i}t
and Garcia-Alfaro, Joaquin",
title="A Tight Integration of Symbolic Execution and Fuzzing (Short Paper)",
booktitle="Foundations and Practice of Security",
year="2022",
publisher="Springer International Publishing",
address="Cham",
pages="303--310",
isbn="978-3-031-08147-7"
}

@inproceedings {cfi_ano,
	author = {Nicholas Carlini and Antonio Barresi and Mathias Payer and David Wagner and Thomas R. Gross},
	title = {{Control-Flow} Bending: On the Effectiveness of {Control-Flow} Integrity},
	booktitle = {24th USENIX Security Symposium (USENIX Security 15)},
	year = {2015},
	isbn = {978-1-939133-11-3},
	address = {Washington, D.C.},
	pages = {161--176},
	url = {https://www.usenix.org/conference/usenixsecurity15/technical-sessions/presentation/carlini},
	publisher = {USENIX Association},
	month = aug
}

@inproceedings{pin,
author = {Luk, Chi-Keung and Cohn, Robert and Muth, Robert and Patil, Harish and Klauser, Artur and Lowney, Geoff and Wallace, Steven and Reddi, Vijay Janapa and Hazelwood, Kim},
title = {Pin: building customized program analysis tools with dynamic instrumentation},
year = {2005},
isbn = {1595930566},
publisher = {Association for Computing Machinery},
address = {New York, NY, USA},
url = {https://doi.org/10.1145/1065010.1065034},
doi = {10.1145/1065010.1065034},
booktitle = {Proceedings of the 2005 ACM SIGPLAN Conference on Programming Language Design and Implementation},
pages = {190–200},
numpages = {11},
location = {Chicago, IL, USA},
series = {PLDI '05}
}

@INPROCEEDINGS{analysis4,
  author={Lin, Yan and Gao, Debin},
  booktitle={2021 IEEE Symposium on Security and Privacy (SP)}, 
  title={When Function Signature Recovery Meets Compiler Optimization}, 
  year={2021},
  volume={},
  number={},
  pages={36-52},
  doi={10.1109/SP40001.2021.00006}}

@misc{tensorflow2015-whitepaper,
  author = {Martín Abadi and others},
  title = {TensorFlow},
  year = {2015},
  howpublished = {\url{https://www.tensorflow.org}},
}

@misc{keras,
  author = {François Chollet and others},
  title = {Keras},
  year = {2015},
  howpublished = {\url{https://keras.io}},
}

@article{dsa_f,
author = {Lattner, Chris and Lenharth, Andrew and Adve, Vikram},
title = {Making context-sensitive points-to analysis with heap cloning practical for the real world},
year = {2007},
issue_date = {June 2007},
publisher = {Association for Computing Machinery},
address = {New York, NY, USA},
volume = {42},
number = {6},
issn = {0362-1340},
url = {https://doi.org/10.1145/1273442.1250766},
doi = {10.1145/1273442.1250766},
journal = {SIGPLAN Not.},
month = jun,
pages = {278–289},
numpages = {12}
}

@inproceedings{cfi1,
author = {Abadi, Mart\'{\i}n and Budiu, Mihai and Erlingsson, \'{U}lfar and Ligatti, Jay},
title = {Control-flow integrity},
year = {2005},
isbn = {1595932267},
publisher = {Association for Computing Machinery},
address = {New York, NY, USA},
url = {https://doi.org/10.1145/1102120.1102165},
doi = {10.1145/1102120.1102165},
booktitle = {Proceedings of the 12th ACM Conference on Computer and Communications Security},
pages = {340–353},
numpages = {14},
location = {Alexandria, VA, USA},
series = {CCS '05}
}

@inproceedings{aict,
author = {Lu, Kangjie and Hu, Hong},
title = {Where Does It Go? Refining Indirect-Call Targets with Multi-Layer Type Analysis},
year = {2019},
isbn = {9781450367479},
publisher = {Association for Computing Machinery},
address = {New York, NY, USA},
url = {https://doi.org/10.1145/3319535.3354244},
doi = {10.1145/3319535.3354244},
booktitle = {Proceedings of the 2019 ACM SIGSAC Conference on Computer and Communications Security},
pages = {1867–1881},
numpages = {15},
location = {London, United Kingdom},
series = {CCS '19}
}

@article{cfi2,
author = {Abadi, Mart\'{\i}n and Budiu, Mihai and Erlingsson, \'{U}lfar and Ligatti, Jay},
title = {Control-flow integrity principles, implementations, and applications},
year = {2009},
issue_date = {October 2009},
publisher = {Association for Computing Machinery},
address = {New York, NY, USA},
volume = {13},
number = {1},
issn = {1094-9224},
url = {https://doi.org/10.1145/1609956.1609960},
doi = {10.1145/1609956.1609960},
journal = {ACM Trans. Inf. Syst. Secur.},
month = nov,
articleno = {4},
numpages = {40}
}

@inproceedings{cfi3,
  title={CFInsight: A Comprehensive Metric for CFI Policies.},
  author={Frassetto, Tommaso and Jauernig, Patrick and Koisser, David and Sadeghi, Ahmad-Reza},
  booktitle={NDSS},
  year={2022}
}

@misc{siamese,
      title={Deep Learning for Answer Sentence Selection}, 
      author={Lei Yu and Karl Moritz Hermann and Phil Blunsom and Stephen Pulman},
      year={2014},
      eprint={1412.1632},
      archivePrefix={arXiv},
      primaryClass={cs.CL},
      url={https://arxiv.org/abs/1412.1632}, 
}

@inproceedings{siamese2,
 author = {Bromley, Jane and Guyon, Isabelle and LeCun, Yann and S\"{a}ckinger, Eduard and Shah, Roopak},
 booktitle = {Advances in Neural Information Processing Systems},
 editor = {J. Cowan and G. Tesauro and J. Alspector},
 pages = {},
 publisher = {Morgan-Kaufmann},
 title = {Signature Verification using a "Siamese" Time Delay Neural Network},
 url = {https://proceedings.neurips.cc/paper_files/paper/1993/file/288cc0ff022877bd3df94bc9360b9c5d-Paper.pdf},
 volume = {6},
 year = {1993}
}

@inproceedings{transformer,
author = {Vaswani, Ashish and Shazeer, Noam and Parmar, Niki and Uszkoreit, Jakob and Jones, Llion and Gomez, Aidan N. and Kaiser, \L{}ukasz and Polosukhin, Illia},
title = {Attention is all you need},
year = {2017},
isbn = {9781510860964},
publisher = {Curran Associates Inc.},
address = {Red Hook, NY, USA},
booktitle = {Proceedings of the 31st International Conference on Neural Information Processing Systems},
pages = {6000–6010},
numpages = {11},
location = {Long Beach, California, USA},
series = {NIPS'17}
}

@article{symexectest,
author = {King, James C.},
title = {Symbolic execution and program testing},
year = {1976},
issue_date = {July 1976},
publisher = {Association for Computing Machinery},
address = {New York, NY, USA},
volume = {19},
number = {7},
issn = {0001-0782},
url = {https://doi.org/10.1145/360248.360252},
doi = {10.1145/360248.360252},
journal = {Commun. ACM},
month = jul,
pages = {385–394},
numpages = {10}
}

@misc{x861,
  title = {System V Application Binary Interface: AMD64 Architecture Processor Supplement},
  author = {{UNIX System Laboratories}},
  year = {2013},
  note = {\url{https://refspecs.linuxfoundation.org/elf/x86_64-abi-0.99.pdf}}
}

@manual{x862,
  title = {Intel 64 and IA-32 Architectures Software Developer’s Manual, Volumes 1–3},
  author = {{Intel Corporation}},
  year = {2021},
  note = {\url{https://www.intel.com/content/www/us/en/developer/articles/technical/intel-sdm.html}}
}

@misc{x32,
  title = {System V Application Binary Interface: Intel386 Architecture Processor Supplement},
  author = {{UNIX System Laboratories}},
  year = {1997},
  note = {\url{https://refspecs.linuxfoundation.org/elf/abi386-4.pdf}}
}

@book{rl,
  title={Reinforcement Learning: An Introduction},
  author={Sutton, Richard S. and Barto, Andrew G.},
  year={2018},
  edition={2nd},
  publisher={MIT Press},
  isbn={978-0262039246},
  url={http://incompleteideas.net/book/the-book-2nd.html}
}

@misc{rl1,
      title={Deep Reinforcement Learning: An Overview}, 
      author={Yuxi Li},
      year={2018},
      eprint={1701.07274},
      archivePrefix={arXiv},
      primaryClass={cs.LG},
      url={https://arxiv.org/abs/1701.07274}, 
}

@ARTICLE{softsignature,
  author={Ullah, Sami and Jin, Wenhui and Oh, Heekuck},
  journal={IEEE Access}, 
  title={Efficient Features for Function Matching in Multi-Architecture Binary Executables}, 
  year={2021},
  volume={9},
  number={},
  pages={104950-104968},
  doi={10.1109/ACCESS.2021.3099429}}

@inproceedings {taint,
author = {Sanchuan Chen and Zhiqiang Lin and Yinqian Zhang},
title = {{SelectiveTaint}: Efficient Data Flow Tracking With Static Binary Rewriting},
booktitle = {30th USENIX Security Symposium (USENIX Security 21)},
year = {2021},
isbn = {978-1-939133-24-3},
pages = {1665--1682},
url = {https://www.usenix.org/conference/usenixsecurity21/presentation/chen-sanchuan},
publisher = {USENIX Association},
month = aug
}

@inproceedings{gnn_using_cfg1, series={CCS ’17},
   title={Neural Network-based Graph Embedding for Cross-Platform Binary Code Similarity Detection},
   url={http://dx.doi.org/10.1145/3133956.3134018},
   DOI={10.1145/3133956.3134018},
   booktitle={Proceedings of the 2017 ACM SIGSAC Conference on Computer and Communications Security},
   publisher={ACM},
   author={Xu, Xiaojun and Liu, Chang and Feng, Qian and Yin, Heng and Song, Le and Song, Dawn},
   year={2017},
   month=oct, pages={363–376},
   collection={CCS ’17} }

@article{gnn_cfg_2, title={Order Matters: Semantic-Aware Neural Networks for Binary Code Similarity Detection}, volume={34}, url={https://ojs.aaai.org/index.php/AAAI/article/view/5466}, DOI={10.1609/aaai.v34i01.5466}, abstractNote={&lt;p&gt;Binary code similarity detection, whose goal is to detect similar binary functions without having access to the source code, is an essential task in computer security. Traditional methods usually use graph matching algorithms, which are slow and inaccurate. Recently, neural network-based approaches have made great achievements. A binary function is first represented as an control-flow graph (CFG) with manually selected block features, and then graph neural network (GNN) is adopted to compute the graph embedding. While these methods are effective and efficient, they could not capture enough semantic information of the binary code. In this paper we propose semantic-aware neural networks to extract the semantic information of the binary code. Specially, we use BERT to pre-train the binary code on one token-level task, one block-level task, and two graph-level tasks. Moreover, we find that the order of the CFG’s nodes is important for graph similarity detection, so we adopt convolutional neural network (CNN) on adjacency matrices to extract the order information. We conduct experiments on two tasks with four datasets. The results demonstrate that our method outperforms the state-of-art models.&lt;/p&gt;}, number={01}, journal={Proceedings of the AAAI Conference on Artificial Intelligence}, author={Yu, Zeping and Cao, Rui and Tang, Qiyi and Nie, Sen and Huang, Junzhou and Wu, Shi}, year={2020}, month={Apr.}, pages={1145-1152} }

@misc{aflpin,
  author       = {van Hauser, THC},
  title        = {{AFL-Pin: Coverage-guided fuzzing with PIN instrumentation}},
  howpublished = {\url{https://github.com/vanhauser-thc/afl-pin}},
  note         = {Accessed: 2025-08-18}
}

@unknown{llm_cfg_1,
author = {Liu, Peipei and Sun, Jian and Chen, Li and Yan, Zhaoteng and Zhang, Peizheng and Sun, Dapeng and Wang, Dawei and Li, Dan},
year = {2025},
month = {03},
pages = {},
title = {Control Flow-Augmented Decompiler based on Large Language Model},
doi = {10.48550/arXiv.2503.07215}
}

@misc{llm_cfg_2,
      title={The CodeInverter Suite: Control-Flow and Data-Mapping Augmented Binary Decompilation with LLMs}, 
      author={Peipei Liu and Jian Sun and Rongkang Sun and Li Chen and Zhaoteng Yan and Peizheng Zhang and Dapeng Sun and Dawei Wang and Xiaoling Zhang and Dan Li},
      year={2025},
      eprint={2503.07215},
      archivePrefix={arXiv},
      primaryClass={cs.SE},
      url={https://arxiv.org/abs/2503.07215}, 
}

@InProceedings{DSA,
    author    = {Chris Lattner and Andrew Lenharth and Vikram Adve},
    title     = "{Making Context-Sensitive Points-to Analysis with Heap Cloning Practical For The Real World}",
    booktitle = "{Proceedings of the 2007 ACM SIGPLAN Conference on Programming Language Design and Implementation (PLDI'07)}",
    address   = {San Diego, California},
    month     = {June},
    year      = {2007}
  }
\appendices


\section{Supporting Details of \sysname{} Components}
\subsection{Recall and F1 Definition} \label{appendix:global_recall}


For consistency, both Global and AICT metrics are defined over $n$ indirect callsites, where $TP_i$, $FP_i$, and $FN_i$ denote the true positives, false positives, and false negatives at callsite~$i$.

\begin{equation*}
\resizebox{\columnwidth}{!}{%
\(
\begin{aligned}
\text{Global\_Precision} &= 
  \frac{\sum_{i=1}^{n} TP_i}{\sum_{i=1}^{n} (TP_i + FP_i)}
&
\text{AICT\_Precision} &= 
  \frac{1}{n} \sum_{i=1}^{n} \frac{TP_i}{TP_i + FP_i}
\\[2pt]
\text{Global\_Recall} &= 
  \frac{\sum_{i=1}^{n} TP_i}{\sum_{i=1}^{n} (TP_i + FN_i)}
&
\text{AICT\_Recall} &= 
  \frac{1}{n} \sum_{i=1}^{n} \frac{TP_i}{TP_i + FN_i}
\\[2pt]
\text{Global\_F1} &= 
  \frac{2 \cdot \text{Global\_Precision} \cdot \text{Global\_Recall}}
       {\text{Global\_Precision} + \text{Global\_Recall}}
&
\text{AICT\_F1} &= 
  \frac{2 \cdot \text{AICT\_Precision} \cdot \text{AICT\_Recall}}
       {\text{AICT\_Precision} + \text{AICT\_Recall}}
\end{aligned}
\)%
}
\end{equation*}

Global metrics aggregate counts across all callsites, while AICT metrics compute per-callsite averages to reflect the analysis quality at the individual callsite level.

\subsection{Heuristics for Argument and Return Value Detection in \layerScoreGen}
\label{appendix:heuristics}

Table~\ref{tab:heuristics} summarizes the heuristic weights used in iScoreGen to filter variables. A variable is retained as a likely argument if its cumulative score exceeds 6.0. For return value detection via \texttt{EAX}, we use analogous rules with a threshold of 2.0.

\begin{table}[tbp]
\centering
\scriptsize
\caption{Heuristic weights for argument and return value detection.}
\label{tab:heuristics}
\begin{tabular}{|l|c|}
\hline
\textbf{Heuristic} & \textbf{Weight} \\
\hline
Positive stack offset & 2.0 \\
Never written  & 1.5 \\
Early use (within first 3 instructions) & 2.0 \\
Used across multiple basic blocks & 1.0 \\
High access frequency  & 1.0 \\
Memory operand with positive offset & 2.0 \\
\hline
\multicolumn{2}{|c|}{\textbf{Return Value (EAX) Weights}} \\
\hline
Written to eax & 1.0 \\
Written in final basic block & 1.5 \\
Assigned a constant value & 1.0 \\
\hline
\end{tabular}
\end{table}

\subsection{\layerScoreGen{} Model Details}
\label{appendix:ml}

The deep neural network used in iScoreGen is a fully connected feedforward model with the architecture shown in Table~\ref{tab:nn_architecture}. We standardize input vectors using \texttt{StandardScaler} before training.

\begin{table}[tbp]
\centering
\tiny
\caption{Neural network architecture and training parameters.}
\label{tab:nn_architecture}
\begin{tabular}{|l|l|}
\hline
\textbf{Layer} & \textbf{Configuration} \\
\hline
Input & Standardized callsite–callee vector \\
Hidden Layer 1 & 1024 neurons, LeakyReLU, Dropout (0.4) \\
Hidden Layer 2 & 512 neurons, LeakyReLU, Dropout (0.4) \\
Hidden Layer 3 & 256 neurons, LeakyReLU, Dropout (0.3) \\
Hidden Layer 4 & 128 neurons, LeakyReLU, Dropout (0.2) \\
Output & 1 neuron, Sigmoid activation \\
\hline
Optimizer & Adam (lr = 0.001) \\
Loss & Binary cross-entropy \\
Early Stopping & Patience = 8 epochs \\
LR Decay & Patience = 4 epochs \\
Batch Size & 256 \\
\hline
\end{tabular}
\end{table}
\subsection{RQ3: Additional CALLEE Evaluation — Address-Taken Function Detection Comparison} \label{appendix:rq3-addr}
Table~\ref{tab:addr_taken_comparison} compares address-taken function detection between \sysname{} and CALLEE. For our compiled binaries, \sysname{} consistently recovers more functions across all benchmarks, while CALLEE misses a sizable fraction of ground-truth callees, especially in larger binaries such as \texttt{gcc}. 
\begin{table}[H]
\centering
\scriptsize
\caption{Comparison of Address-Taken Function Detection.}
\label{tab:addr_taken_comparison}
\begin{tabular}{|l|c|c|c|c|}
\hline
\textbf{Binary} & \textbf{\sysname{}} & \textbf{CALLEE} & \textbf{GT Missed} & \textbf{\% Missed} \\
\hline
gobmk       & 1787  & 1757  & 29   & 8   \\
perlbench   & 720   & 689   & 9    & 3   \\
gcc         & 1207  & 414   & 138  & 33.4  \\
\hline
\end{tabular}
\caption*{\scriptsize GT Missed: ground truth callees missed by CALLEE. \% Missed is over total GT callees.}
\end{table}



\begin{figure}[t]
  \centering
  \begin{subfigure}[t]{0.30\columnwidth}
    \includegraphics[width=\linewidth]{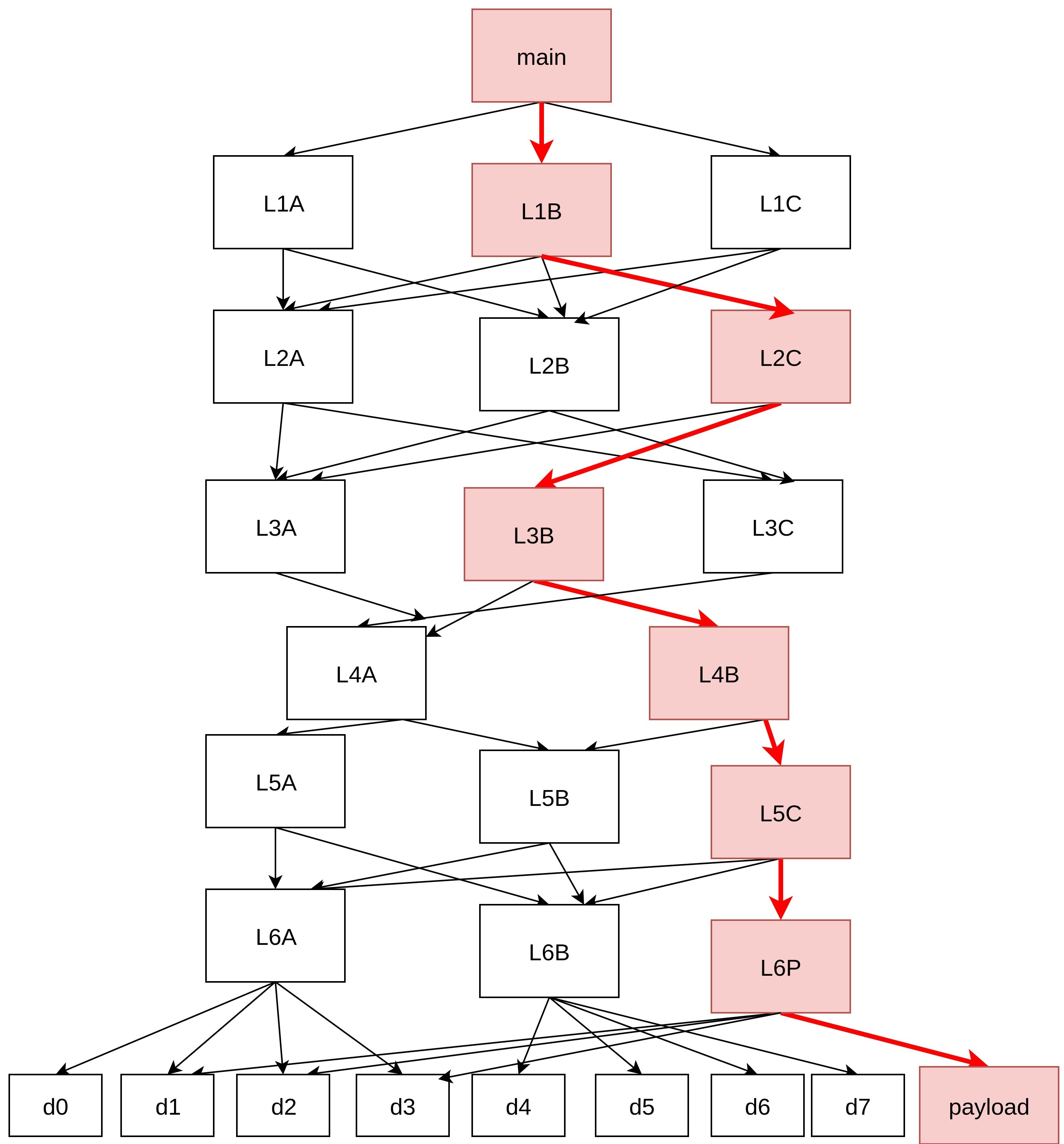}
    \caption{True Call Graph}
    \label{fig:trueCG}
  \end{subfigure}
  \hfill
  \begin{subfigure}[t]{0.30\columnwidth}
    \includegraphics[width=\linewidth]{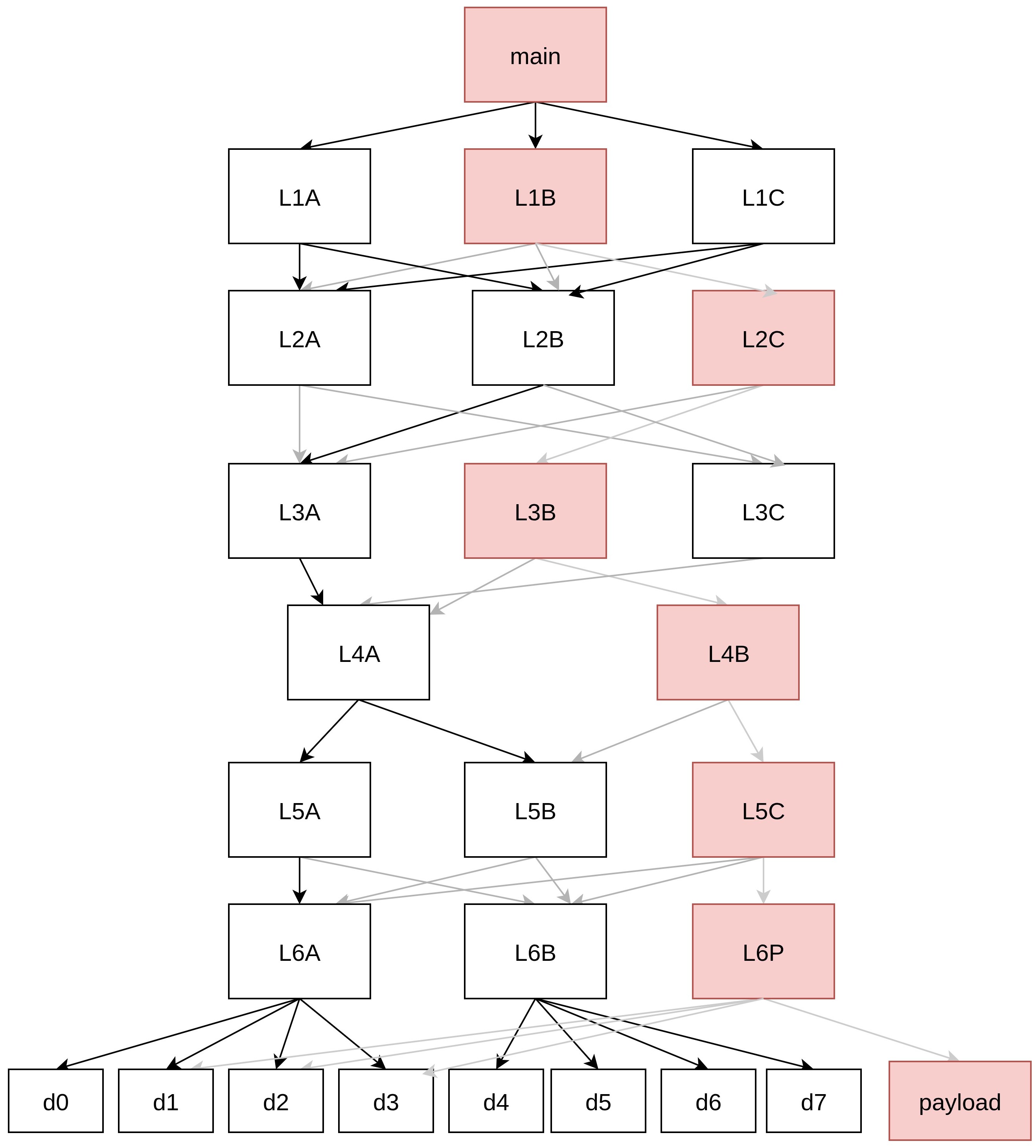}
    \caption{IDA Generated Call Graph}
    \label{fig:IDACG}
  \end{subfigure}
  \hfill
  \begin{subfigure}[t]{0.30\columnwidth}
    \includegraphics[width=\linewidth]{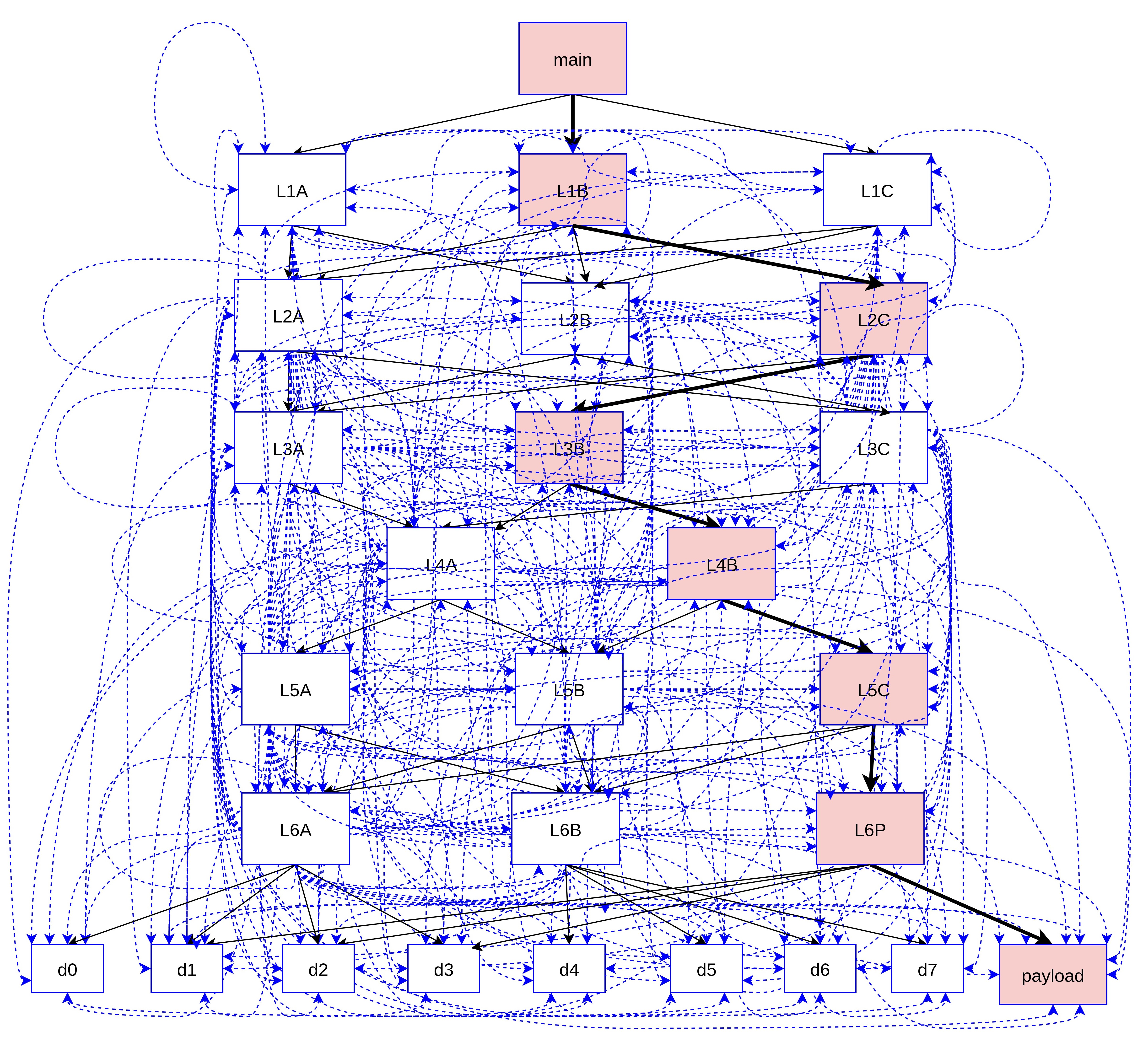}
    \caption{BPA Generated Call Graph}
    \label{fig:BPACG}
  \end{subfigure}

  \vspace{1mm} 

  \begin{subfigure}[t]{0.30\columnwidth}
    \includegraphics[width=\linewidth]{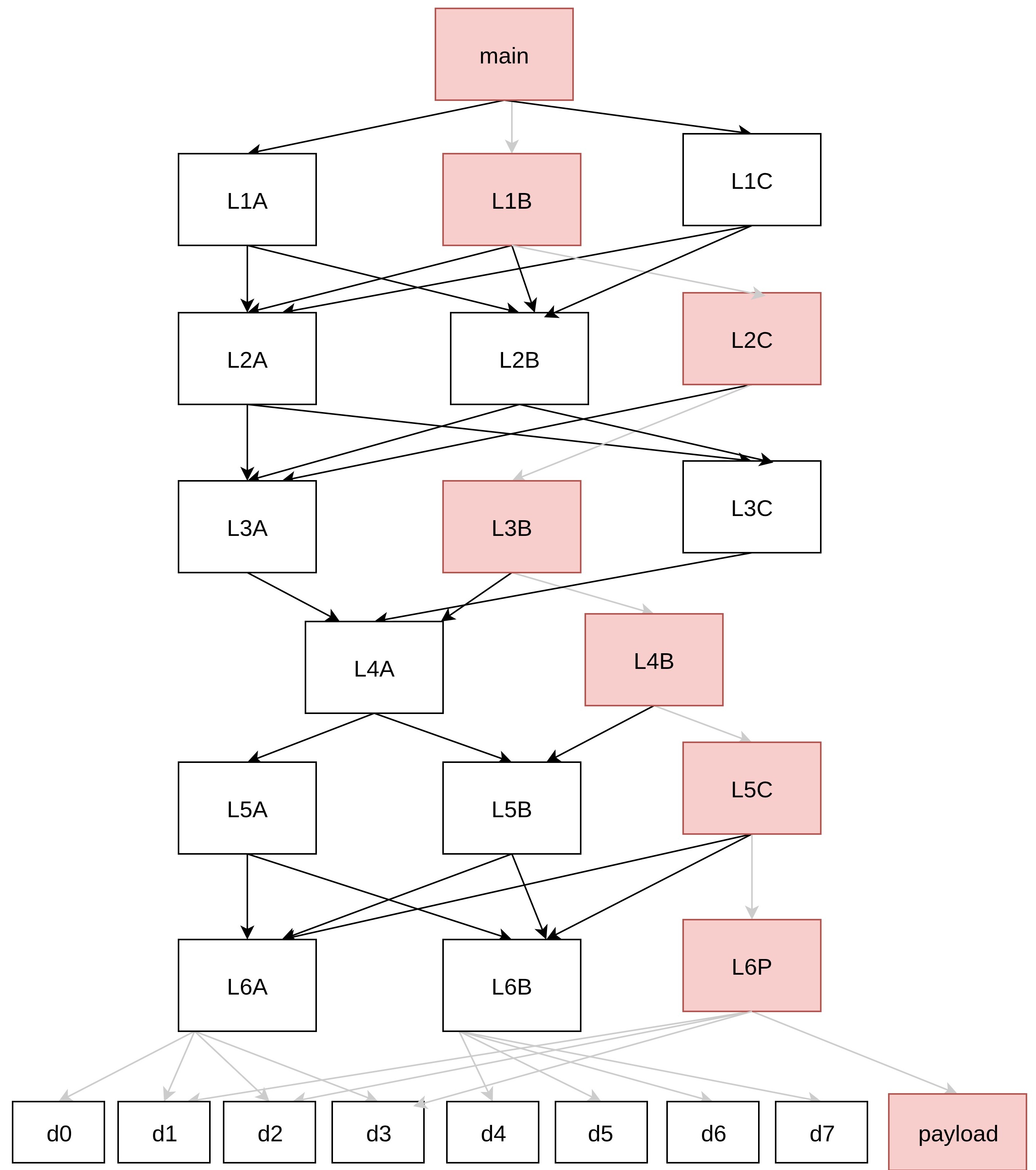}
    \caption{p-IndirectCFG at \texttt{Threshold>=0.7}}
    \label{fig:us_7}
  \end{subfigure}
  \hfill
  \begin{subfigure}[t]{0.30\columnwidth}
    \includegraphics[width=\linewidth]{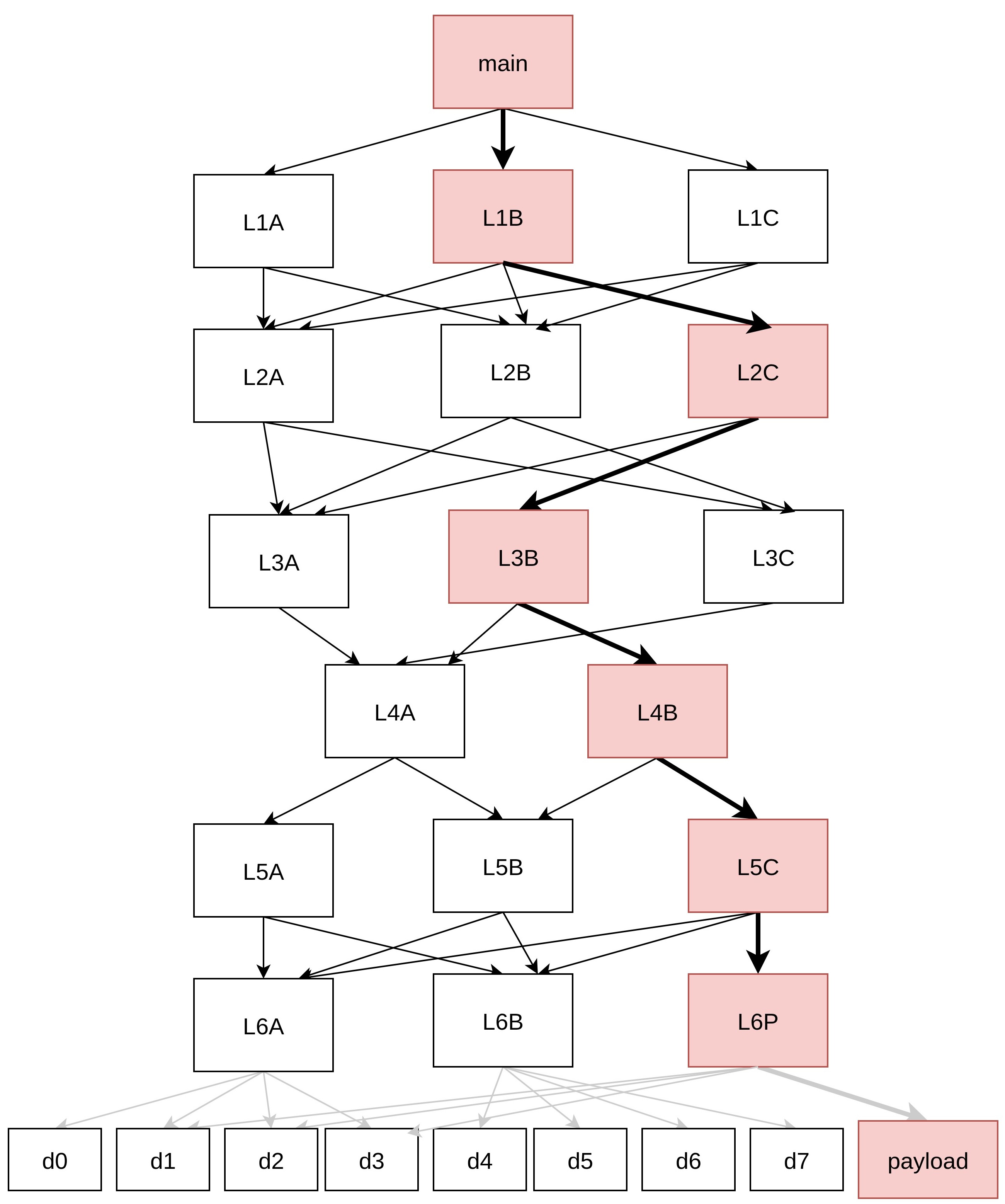}
    \caption{p-IndirectCFG at \texttt{Threshold>=0.6}}
    \label{fig:us_6}
  \end{subfigure}
  \hfill
  \begin{subfigure}[t]{0.30\columnwidth}
    \includegraphics[width=\linewidth]{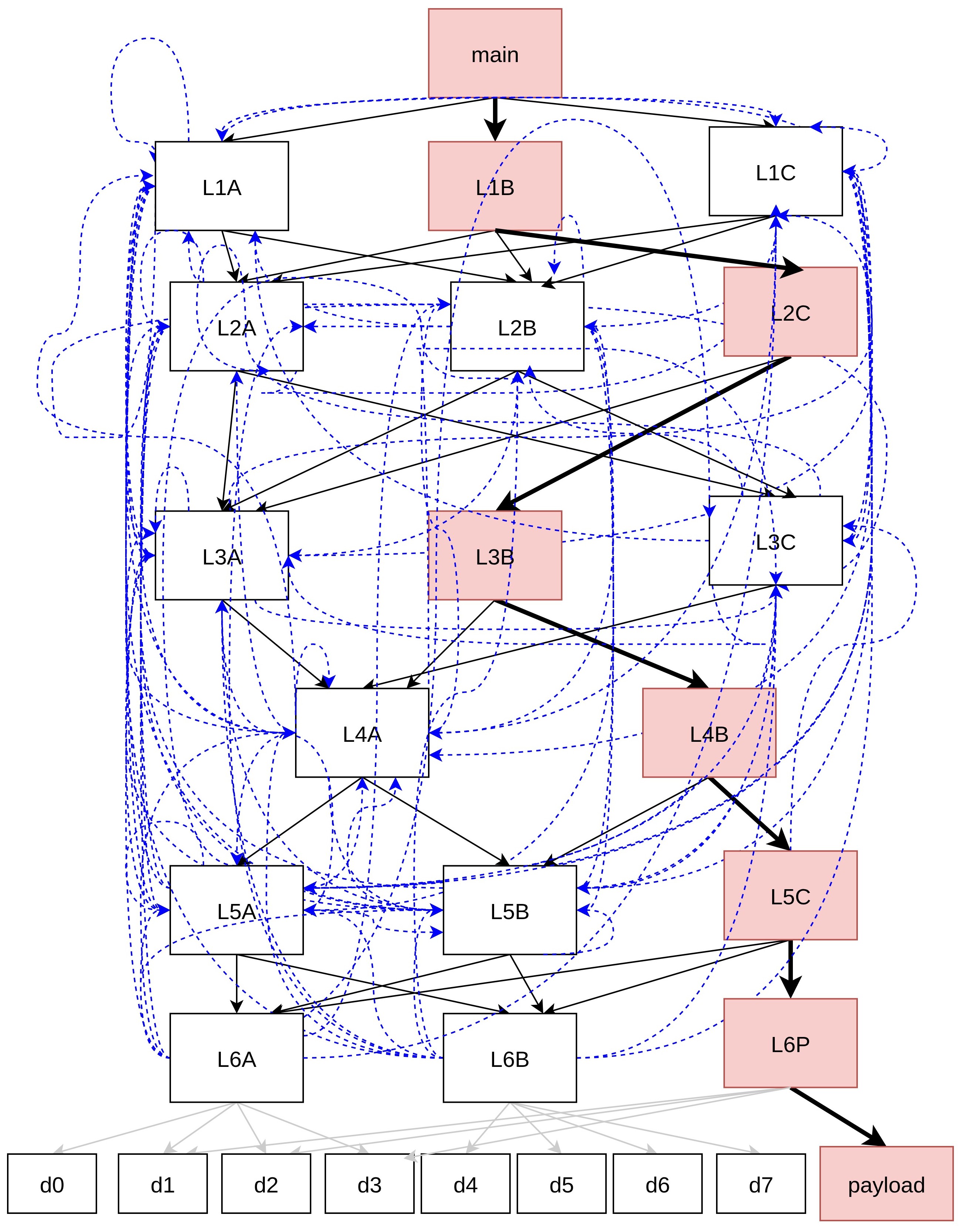}
    \caption{p-IndirectCFG at \texttt{Threshold>=0.57}}
    \label{fig:us_57}
  \end{subfigure}

  \caption{PoC Binary Call Graphs by Different Tools. (Black: tool-detected; Blue dotted: false positives; Gray: ground truth edges). Color coding applies to all graphs except the True Call Graph.} 
  \label{fig:recall-poc}
\end{figure}

\subsection{Case Study --- Why p-IndirectCFG Matters in Real-World RE} \label{appendix:case_study}
The ideal outcome for indirect call resolution is a complete and precise CFG, yet no existing tool---academic or commercial---achieves both. Reverse engineers are left juggling partial CFGs from multiple tools (IDA, Ghidra, academic prototypes) or resorting to dynamic tracing. Sound tools like BPA ensure recall but overwhelm analysts with false positives, undermining both manual inspection and automated reasoning.
p-IndirectCFG bridges this gap with a flexible, confidence-annotated CFG supporting both precision-critical and recall-critical tasks. By adjusting its pruning threshold, it balances false positives against coverage. We demonstrate this via two contrasting scenarios:

\begin{lstlisting}[float, floatplacement=htbp,style=csecurity,caption={PoC for recall-sensitive case study},label={lst:recall_poc}]
/* 
 * Only main->L1B->L2C->L3B->L4B->L5C->L6P->payload uses pointer aliasing:
 * template -> local[] -> p -> pp -> use. All other paths use simple index
 * selection from a local copy of a function-pointer array.
 */

typedef void (*fp_t)(const char *);
void payload(const char *s) { puts("CRITICAL_PAYLOAD"); }
void d1(const char *s) { (void)s; } // Decoys (d0 to d7 identical)

// --- Global memory ---
static const fp_t L1_TABLE[] = { L1A, L1B, L1C }; // Only L1B reaches payload

// --- L6P: leaf with aliasing and payload ---
static const fp_t L6P_TEMPLATE[] = { d1, d2, payload, d3 };
// Each function (e.g., L6P, L5P, L4A, etc.) has its own global template

void L6P(const char *arg) {
fp_t local[L6P_TEMPLATE_SIZE]; 
for (int i = 0; i < L6P_TEMPLATE_SIZE; ++i) local[i] = L6P_TEMPLATE[i];
fp_t *p = local; fp_t **pp = &p; fp_t *use = *pp;
unsigned idx = pick_index(arg, L6P_TEMPLATE_SIZE - 1);  
use[idx]("via L6P");  // May call payload
}

// --- L5C: contains L6P in template ---
/* L5C() rematerializes and aliases its template {L6A, L6B, , L6P}, 
   then dispatches like L6P; may call L6P. */

// ----- Intermediate Layers -----
// L1B() -> selects from { L2A, L2C, L2B }; only L2C leads to payload chain
// L2C() -> selects from { L3B, ... }; only L3B reaches L4B
// L3B() -> selects from { L4A, L4B}; only L4B reaches L5C
// L4B() -> selects from { L5B, L5C }; only L5C reaches L6P


// --- Main driver ---
int main(int argc, char **argv) {
  const char *arg = (argc > 1) ? argv[1] : "foo";
  unsigned idx = pick_index(arg, L1_TABLE_SIZE - 1); 
  fp_t entry = L1_TABLE[idx];
  entry(arg); // -> may enter alias-based payload chain, selectes from {L1A, L1B, L1C}, Only L1B can transitively reach payload.
}
\end{lstlisting}

\textbf{Recall-Sensitive Scenario:} Our PoC binary mimics real malware that hides payloads behind multi-layered pointer aliasing, memory–code dependencies, and inter-procedural indirection. Out of 232 paths, only one reaches the payload. The PoC uses both index-based dispatch and alias-driven function pointer dereference to simulate realistic obfuscation (see Listing~\ref{lst:recall_poc}). Only the path \texttt{main → L1B → L2C → L3B → L4B → L5C → L6P → payload} reaches the hidden function (Figure~\ref{fig:trueCG}). Ghidra fails entirely; IDA partially resolves calls but misses the payload (Figure~\ref{fig:IDACG}). BPA achieves 100\% recall but with an AICT of 16.9---at first glance, 16.9 average targets
per callsite may seem small in absolute terms, but for large
programs this quickly scales to hundreds or thousands of
extra edges (Figure~\ref{fig:BPACG}). The spurious edges lead to structural distortions, e.g., the payload appears reachable via false paths, and static analyses like symbolic execution are overwhelmed by unrealizable paths.

In contrast, p-IndirectCFG progressively sharpens recall-precision tradeoffs: at a high threshold (Figure~\ref{fig:us_7}), only high-confidence edges are retained---no FPs, but incomplete path. Lowering the threshold (Figure~\ref{fig:us_6}) recovers most of the malicious chain with 0 FPs, exceeding IDA/Ghidra. Further relaxation (Figure~\ref{fig:us_57}) captures the full payload path while still reducing noise compared to BPA.
Such flexibility enables automated traversal strategies based on edge confidence, useful for manual or even agent-based malware analysis. This illustrates the necessity of recall sensitivity in security-critical RE. We also tested CALLEE. As discussed in Section~\ref{sec:rq3}, we couldn’t reproduce its results. Our re-engineered CALLEE CFGs were more imprecise than BPA's, likely due to slicing limits in presence of argument similarity and deep memory aliasing. The resulting over-approximation renders them impractical for RE workflows.

\textbf{Precision-Sensitive Scenario:} Increased false positives severely degrade downstream analysis. A single spurious inter-procedural edge can merge regions, create non-existent loops, and inflate fixpoint iterations. Symbolic execution suffers from path explosion and solver timeouts; dominator trees and SSA form gain artificial joins, dominance and structure are distorted, and variables/types collapse to ``top'' or unknown. Inter-procedural analyses face context blow-up as each extra callee multiplies call strings and summary unions. Slicing and taint analysis over-approximate, pulling in irrelevant callers and sinks. Noisy CFGs also harm ML, causing models to adapt to bad edges and underperform on clean graphs. For general-purpose analysis, it is often better to keep false positives low—even at some recall cost—so long as the CFG remains faithful to the program’s intended semantics.

We illustrate this with two SPEC benchmarks where all RE tools miss most indirect calls and static tools overwhelm the CFG with false positives, particularly as program size grows (from \texttt{h264ref} to \texttt{gobmk}).

In \texttt{h264ref}-O0, two global structures govern indirect dispatch. A \texttt{DataPartition} field \texttt{writeSyntaxElement} selects between UVLC and CABAC encoders (first indirection), while a \texttt{SyntaxElement} field (\texttt{mapping} for UVLC, \texttt{writing} for CABAC) drives a second indirection inside each encoder (see Listing~\ref{lst:h264ref-two-step-indirection}). The UVLC path never calls CABAC routines and vice versa, so any CFG edge crossing between these families is semantically invalid. RE tools fail to resolve many targets here, and BPA introduces numerous false positives. In particular, BPA creates a directed cycle where \texttt{writeSyntaxElement\_CABAC} appears to call \texttt{writeSyntaxElement\_UVLC} and then return to itself, plus self-loops on both routines, at AICT $= 5.7$ with 100\% recall. \sysname{}'s refinement first reduces AICT to 3.4 (spurious paths still present), then to 1.68 at 95.4\% recall, eliminating the cycle and one self-loop. Restricting to high-confidence edges removes all of these false paths at 80\% recall, yielding a CFG faithful enough for precision-critical analysis. By contrast, CALLEE and AttnCall produce AICT values $\ge 20$ on this case, exceeding even BPA’s false-positive rate and introducing more spurious paths.

The problem worsens on larger binaries such as \texttt{gobmk}-O0. Here, two unrelated indirection families coexist: pattern evaluation, where \texttt{check\_pattern\_hard} dispatches only via \texttt{pattern->autohelper}/\texttt{pattern->helper}, and GTP command handling, where \texttt{gtp\_main\_loop} dispatches only via \texttt{commands[i].function} (see Listing~\ref{lst:gobmk-two-indirections}). There is a strict semantic boundary between these families; any edge from pattern callbacks into GTP handlers (or vice versa) is spurious. BPA’s coarse global memory model nonetheless mixes them, yielding 3{,}554 false cycles (maximum loop length 19) versus just 115 legitimate cycles—a \textbf{2{,}990\%} increase—at AICT $= 884.6$ with 100\% recall. This distorts the CFG and induces path explosion in downstream analyses. Using p-IndirectCFG, \sysname{} raises the pruning threshold to remove all spurious cycles and cuts AICT by 50\% while still maintaining 95.5\% recall, restoring semantic integrity without polluting downstream workflows.

\begin{lstlisting}[float,style=csecurity,
caption={h264ref two-step indirection: First, select UVLC or CABAC encoding via \texttt{DataPartition->writeSyntaxElement}; Second, indirect via SyntaxElement.mapping/SyntaxElement.writing}.,
label={lst:h264ref-two-step-indirection}]
/* -- §\hlc{Data Structures related to first indirection}§ -- */
typedef struct datapartition {
  /* ... other fields ... */
  int (*writeSyntaxElement)(SyntaxElement*, struct datapartition*);  // first indirection (UVLC or CABAC)
} DataPartition;

/* -- §\hlc{Data Structures related to second indirection}§ -- */
typedef struct syntaxelement {
  int  type, value1, value2, len, inf, context;
  void (*mapping)(int, int, int*, int*);                              // used in second indirection (UVLC)
  void (*writing)(struct syntaxelement*, EncodingEnvironmentPtr);    // used in second indirection (CABAC)
} SyntaxElement;

/* -- §\hlc{First indirection setup (0=UVLC, 1=CABAC) }§-- */
static void init_slice(int start_mb_addr) {
  /* ... other steps ... */
  dataPart->writeSyntaxElement =
      (input->symbol_mode == UVLC) ? writeSyntaxElement_UVLC
                                   : writeSyntaxElement_CABAC;
}

/* -- §\hlc{Second-indirection setup and invoke first indirection callsite}§ -- */
for (int k = 0, level = 1; k <= 16 && level != 0; k++) {
  /* ... other steps ... */

  if (input->symbol_mode == UVLC)
    currSE->mapping = levrun_linfo_inter; // UVLC: pick "linfo" mapper
  else
    currSE->writing = writeRunLevel_CABAC;  // CABAC: pick writer

  /* ... other steps ... */
  dataPart->writeSyntaxElement(currSE, dataPart); // first indirection
}

/* -- §\hlc{Second indirection callsite: CABAC}§ -- */
int writeSyntaxElement_CABAC(SyntaxElement *se, DataPartition *dp) {
  /* ... CABAC encoding steps ... */
  se->writing(se, _);   // second indirection

}

/* -- §\hlc{Second indirection callsite: UVLC}§ -- */
int writeSyntaxElement_UVLC(SyntaxElement *se, DataPartition *dp) {
  se->mapping(se->value1, se->value2, &se->len, &se->inf); // second indirection
  /* ... UVLC encoding steps ... */
}
\end{lstlisting}

\begin{lstlisting}[ style=csecurity,
caption={gobmk: minimal two indirection explanation. First, Pattern evaluation calls only \texttt{pattern->autohelper}/\texttt{pattern->helper}. Second, GTP loop dispatches only via \texttt{commands[i].function}.},
label={lst:gobmk-two-indirections}]
/* -- §\hlc{Data Structures related to first indirection}§ -- */
struct pattern {
  int    autohelper_flag;
  double constraint_cost;
  int  (*autohelper)(int ll, int move, int color, int what); /* indirection */
  int  (*helper)(struct pattern *, int rotation,
				     int move, int color);                       /* indirection */
};

/* -- §\hlc{Data Structures related to second indirection}§ -- */
/* -- Types for GTP command dispatch -- */
struct gtp_command {
  const char *name;
  int (*function)(char *args); 
};

/* Minimal command table (illustrative subset) */
static int gtp_play(char *p);
static int gtp_genmove(char *p);
static struct gtp_command commands[] = {
  {"play",    gtp_play},
  {"genmove", gtp_genmove},
  /* ... many other commands ... */
  {NULL,      NULL}
};

/* -- §\hlc{Indirect call site 1: pattern evaluation}§  -- */
static int
check_pattern_hard(int move, int color, struct pattern *pattern, int ll)
{
  /* indirection via pattern->autohelper */
  if ((pattern->autohelper_flag & HAVE_CONSTRAINT)
      && pattern->constraint_cost < 0.45) {
    if (!pattern->autohelper(ll, move, color, 0))
      return 0;
  }
  return 1;
}

/* -- §\hlc{Indirect call site 2: GTP main loop dispatch}§ -- */
void
gtp_main_loop(struct gtp_command commands[], FILE *gtp_input)
{
  char line[GTP_BUFSIZE], command[GTP_BUFSIZE]; char *args;
  while (fgets(line, GTP_BUFSIZE, gtp_input)) {
    /* ... preprocess & parse line into command + args ... */
    /* Example (simplified): */
    if (sscanf(line, " %s", command) < 1) continue;
    args = strchr(line, ' '); if (args) while (*args==' ') ++args;

    /* indirection via commands[i].function */
    for (int i = 0; commands[i].name; i++) {
      if (strcmp(command, commands[i].name) == 0) {
        (commands[i].function)(args);
        break;
      }
    }
  }
}
\end{lstlisting}



%



\end{document}